# Anomalous mass renormalization in two-dimensional electron systems on ultra-wide bandgap semiconductor β-$Ga_2O_3$

**Ryu Yukawa,[1,*] Hiroshi Mizuseki,[2] Suryo Santoso Putro,[3] Yé-Jin L. Lee,[1,4] Yuuki Masutake,[3] Hinako Telengut,[3] Boxuan Li,[3] Hajime Yamamoto,[3] Tadashi Abukawa,[1,3] Junya Yoshida,[1] Vladimir V. Kochurikhin,[5,6] Taketoshi Tomida,[5,6,7] Masanori Kitahara,[5,6,7] Takahiko Horiai,[6,8] Akira Yoshikawa,[5,6,7,8] Nobuhiko Sarukura,[8,9] Noriko Chikumoto,[9] Toshihiko Shimizu,[8,9] Marilou Cadatal-Raduban,[9,10] Yoshiyuki Kawazoe,[8,11,12] Ryuhei Kohno,[1] Hiroshi Kumigashira,[3] Takuto Nakamura,[13,14] Tatsuhiko Kanda,[15] Akira Yasui,[15] Akihiro Koide,[15] Miho Kitamura,[16] Hideaki Iwasawa,[16,17,18] Koji Horiba,[16] Kenichi Ozawa[19]**

[1]International Center for Synchrotron Radiation Innovation Smart (SRIS), Tohoku University, Sendai 980-8572, Japan

[2]Korea Institute of Science and Technology (KIST), Seoul 02792, Republic of Korea

[3]Institute of Multidisciplinary Research for Advanced Materials (IMRAM), Tohoku University, Sendai 980-8577, Japan

[4]Faculté des Sciences et Ingénierie, Sorbonne Université, F-75005 Paris, France

[5]Institute for Materials Research (IMR), Tohoku University, Sendai 980-8577, Japan

[6]C&A Corporation, Sendai 980-0811, Japan

[7]FOX Corporation, Sendai 982-0024, Japan

[8]New Industry Creation Hatchery Center (NICHe), Tohoku University, Sendai 980-8579, Japan

[9]Institute of Laser Engineering, Osaka University, 2-6 Yamadaoka, Suita 565-0871, Japan

[10]Unitec Institute of Technology, 139 Carrington Road, Mount Albert, Auckland 1025, New Zealand
[11]Department of Physics, SRM University-AP, Amaravati 522 240, Andhra Pradesh, India
[12]School of Physics, Institute of Science and Center of Excellence in Advanced Functional Materials, Suranaree University of Technology, Nakhon Ratchasima 30000, Thailand
[13]Graduate School of Frontier Biosciences, Osaka University, Suita 565-0871, Japan
[14]Department of Physics, Graduate School of Science, Osaka University, Toyonaka 560-0043, Japan
[15]Japan Synchrotron Radiation Research Institute (JASRI), Sayo 679-5198, Japan
[16]NanoTerasu Center, National Institutes for Quantum Science and Technology (QST), Sendai 980-8572, Japan
[17]Synchrotron Radiation Research Center, National Institutes for Quantum Science and Technology (QST), Sayo 679-5148, Japan
[18]Research Institute for Synchrotron Radiation Science, Hiroshima University, Higashi-Hiroshima 739-0046, Japan
[19]Institute of Materials Structure Science, High Energy Accelerator Research Organization (KEK), Tsukuba 305-0801, Japan

* e-mail: r.yukawa@tohoku.ac.jp

**Two-dimensional electron systems (2DESs) at wide-bandgap semiconductor surfaces and interfaces are fertile ground for emergent quantum phenomena. Here, we use micro-focused angle-resolved photoemission spectroscopy (ARPES) to resolve the momentum-space electronic structure of a surface 2DES in β-$Ga_2O_3$, an ultra-wide bandgap semiconductor with a band gap of 4.7 eV. *In situ* electron doping of cleaved, high-purity β-$Ga_2O_3$ single crystals produces an essentially circular Fermi surface. With the self-consistent Poisson–Schrödinger model, the 2D electrons are estimated to be confined within approximately 1.2 nm beneath the surface, with an internal electric field of approximately 18 MV cm$^{-1}$. Our analysis based on ARPES measurements reveals nearly a twofold increase in the electron mass at high sheet carrier densities—a trend that deviates from the conventional behavior reported for wide-bandgap semiconductors—indicating mass renormalization beyond the conventional screening-based picture. Our results provide quantitative momentum-space benchmarks for the Fermi-surface geometry and density-dependent quasiparticle dispersion of the β-$Ga_2O_3$ 2DES, opening a route to studying density-tunable interfacial quantum states in the UWBG regime.**

Two-dimensional electron systems (2DESs) at wide-bandgap semiconductor surfaces and interfaces are crucial interfacial conduction channels for power devices[1] and serve as fertile ground for emergent quantum phenomena—including the fractional quantum Hall effect[2], interfacial superconductivity[3,4], and pronounced spin splitting[5,6]—whose characteristics depend sensitively on sheet carrier density. Recent angle-resolved photoemission spectroscopy (ARPES) studies on 2DESs in wide-bandgap semiconductors have shown a carrier-density-driven crossover from polaronic behavior to a Fermi liquid, accompanied by a reduction in the effective mass[7–9]. Therefore, sheet-carrier-density-dependent ARPES measurements are required to identify and explore novel quantum phenomena in 2DESs.

Although ARPES has been extensively applied to conduction-band-derived 2DESs in wide-gap materials[7–12], previous studies have focused primarily on hosts with band gaps of approximately 3.6 eV or less, leaving such states in UWBG semiconductors ($E_g$ > 4 eV) largely unexplored. Owing to their large band gaps, UWBG materials exhibit high critical breakdown fields and are therefore expected to host new quantum phenomena and to play key roles in the design of next-generation power electronics[13]. Among UWBG semiconductors, β-$Ga_2O_3$ is widely regarded as a promising candidate for power electronics[14–16]: its 4.7 eV band gap[17] supports a critical breakdown field of up to 8 MV $cm^{-1}$ (Ref. 18), yielding a Baliga figure of merit about four times that of GaN and nearly an order of magnitude greater than 4H-SiC. Its high thermal stability[19,20] further strengthens its prospects for high-temperature operation. Despite extensive electrical characterization of β-$Ga_2O_3$, key interfacial parameters—the effective mass and the intrinsic in-plane isotropy—remain unresolved. Electrical-transport studies show divergent pictures of conduction-band anisotropy in β-$Ga_2O_3$: a four-probe study on single crystals reported more than a 17-fold higher conductivity along the *b* axis than along the *c* axis[21] (crystallographic axes are defined in Fig. 1**a**), whereas recent conductivity studies show nearly isotropic transport[22,23], motivating a direct, momentum-resolved determination of the band structure. Despite this importance, the momentum-resolved electronic structure of a 2DES on β-$Ga_2O_3$ remains largely unexplored, primarily because of the difficulty in preparing pristine surfaces and flat cleavage planes large enough for ARPES measurements.

In this study, we elucidate the electronic structure of the 2DES on the cleaved surfaces of high-purity β-$Ga_2O_3$ single crystals by performing micro-focused ARPES. We synthesized β-$Ga_2O_3$ single crystals using the oxide-crystal growth from cold crucible (OCCC) method[24], in which a high-purity single crystal is synthesized without an iridium crucible; the resulting crystal withstands photon-induced damage during ARPES with synchrotron light and enables high-precision electronic structure measurements. Electron doping produces an isotropic, circular Fermi surface, while self-consistent Poisson–Schrödinger modeling estimates that the electrons are confined within ~1.2 nm beneath the surface and experience an interfacial field of ~18 MV $cm^{-1}$. Strikingly, as the sheet carrier density increases, the effective mass grows from a value close to the calculated bare band mass to nearly twice that value—opposite to the trend observed for other oxide 2DESs—thereby revealing a distinct renormalization regime in β-$Ga_2O_3$. These results establish a quantitative, momentum-resolved baseline for the β-$Ga_2O_3$ interfaces and reveal a distinct density-dependent renormalization regime in the UWBG semiconductor.

## Results

For all ARPES and optical measurements, we used high-purity β-$Ga_2O_3$ single crystals (Fig. 1**b**) grown by the crucible-free OCCC method[24]. The optical band gap, determined from the Tauc method where plots of $(\alpha h\nu)^2$ versus photon energy are extrapolated linearly, yields $E_\mathrm{g} = 4.7$ eV, a value consistent with earlier work[17]. ARPES measurements were conducted on the surface of the $bc$ plane (Fig. 1**a**). We define the $x, y$, and $z$ axes as the directions parallel to the $b$-axis (also parallel to the analyzer slit), the $c$-axis, and the surface normal, respectively. Because the conduction-band minimum is located at the bulk Brillouin-zone center (Γ; see Supplementary Fig. S10), we acquired ARPES spectra in the vicinity of $\bar{\Gamma}$, the surface-projected image of Γ in the two-dimensional Brillouin zone.

On the electron-doped surface, ARPES measurements reveal a clear parabolic band with its bottom located at the binding energy ($E_\mathrm{B}$) of ~500 meV (Fig. 1**d**). MDC analysis of the Fermi-level crossings yields a Fermi wavenumber of $k_\mathrm{F} = 0.25$ Å$^{-1}$. A momentum map integrated over $E_\mathrm{F} \pm 20$ meV shows a nearly circular contour in the $k_x$–$k_y$ plane (Fig. 1**e**). A detailed angular analysis revealed no statistically significant difference between the $k_x$ and $k_y$ directions at the 95% confidence level (see Supplementary Fig. S9 for details). In contrast, the intensity map in the $k_x$–$k_z$ plane exhibits no dispersion along the $k_z$ direction (Fig. 1**f**), which confirms the two-dimensional nature of these electronic states. Although a slight shrinkage of the Fermi contour is observed for $k_z > 6$ Å$^{-1}$, most probably due to the partial desorption of alkali atoms (Cs) during the extended measurement, the absence of dispersion over at least one full Brillouin zone in $k_z$ (~1.1 Å$^{-1}$) is strong evidence for a 2DES. These results therefore establish the formation of a surface 2DES on β-$Ga_2O_3$, with no resolvable in-plane anisotropy within the present experimental uncertainty. Using $k_\mathrm{F}$ of 0.25 Å$^{-1}$ and assuming a spin-degenerate circular Fermi contour, the sheet carrier density is estimated to be $n_\mathrm{2D} = \pi k_\mathrm{F}^2/(2\pi^2) = 1.0 \times 10^{14}$ cm$^{-2}$.

To account for these observations, we compared the ARPES results with density functional theory (DFT) calculations (Fig. 2). The projected density of states (PDOS) in Fig. 2**a** shows that the conduction band minimum is derived mainly from Ga 4$s$ and O 2$s$ orbitals. Consistent with this $s$-orbital character, the calculated in-plane dispersion is virtually orientation-independent: at $E$ - $E_{CBM}$ = 0.5 eV, the momentum difference between the $k_x$ and $k_y$ directions is smaller than 5%. These calculations support the isotropic Fermi surface observed in ARPES (Fig. 1**e**) and confirm that the metallic state originates from conduction-band subbands quantized in the near-surface potential well.

Transport-relevant coherence was quantified from the MDC at $E_F$. A full width at half maximum $\Delta k_{\mathrm{F}}$, obtained from a double-Lorentzian fit (Fig. 1**d**), yields a quasiparticle coherence length $l = 1/\Delta k_F = 6.7 \pm 0.3$ Å along the $k_x$ direction ($5.5 \pm 0.9$ Å along the $k_y$ direction). Because $l$ exceeds the monoclinic $b$-axis lattice constant of 3.04 Å (Ref. 25), the system lies safely beyond the Ioffe–Regel limit[26], indicating coherent in-plane electron propagation. By contrast, self-consistent Poisson–Schrödinger simulations estimate the strongly confined nature of the 2DES along the surface normal direction (Figs. 2**c** and 2**d**); 89% of electrons in the first subband are confined within 1.2 nm, and the carrier-density profile peaks at 0.6 nm below the surface, reaching as high as $1.1 \times 10^{21}$ $cm^{-3}$. Therefore, within this model, the 2DES is estimated to be tightly confined to the immediate surface region along the surface-normal direction. Independently, the ARPES-derived in-plane coherence length exceeds two unit cells along the $b$ direction, supporting the formation of a spatially confined yet laterally coherent 2DES on the UWBG semiconductor β-$Ga_2O_3$ (Fig. 2**e**).

Although the isotropy of the metallic states is well-explained by the DFT calculations, a clear distinction exists in the effective mass of the metallic states. A quadratic fit to the DFT dispersion (Fig. 2**b**) over $0 < E - E_{CBM} < 0.5$ eV yields an effective mass $m_{DFT} = 0.26\ m_e$ ($m_e$ is the mass of a free electron) along the $k_x$ direction, whereas experimental dispersion yields $m^* = 0.48 \pm 0.02\ m_e$ from the quadratic fit to the MDCs, almost twice the value obtained from the calculations. This pronounced enhancement relative to the calculated band mass indicates that the high-density 2DES is not captured by a simple single-particle picture.

To track the evolution of the mass renormalization (Figs. 3**a-d**), we conducted ARPES measurements at various $n_{2D}$ values. At low $n_{2D}$ ($< 2 \times 10^{12}$ $cm^{-2}$) (Fig. 3**a**), no clear quasiparticle dispersion is resolved within the experimental resolution. Instead, a shallow dip-like feature near $E_B$ = 70 meV (marked by arrows) and a long-tailed structure at the higher-binding-energy side are visible. Analogous to the tail structures observed in β-$Ga_2O_3$ (Ref. 27) and other oxide semiconductors[7,28,29] at low carrier densities, we attribute these features to multiple-phonon loss structures that are described by a Franck-Condon model, in which coherent electrons near $E_F$ lose their energy by excitation of longitudinal optical (LO) phonons. Indeed, since β-$Ga_2O_3$ is an ionic semiconductor composed of $Ga^{3+}$ and $O^{2-}$, conducting electrons in the crystal are known to couple strongly to LO phonons[30,31]. As $n_{2D}$ increases, the dip structure is no longer observable at $7.6 \times 10^{12}$ $cm^{-2}$ (Fig. 3**b**), and clear dispersive features become evident at higher $n_{2D}$ (Figs. 3**c** and 3**d**). Soft-X-ray ARPES of Ce-doped $CaMnO_3$ has likewise demonstrated the coexistence of weakly coupled light carriers and strongly coupled polarons, highlighting the sensitivity of the low-energy spectral weight to electron–phonon coupling[32]. The observed spectral evolution is therefore consistent with a crossover from a polaronic regime at low carrier density to a more coherent quasiparticle regime at higher carrier density[7,8], although its microscopic origin cannot be uniquely determined from the present data. At $2.9 \times 10^{13}$ $cm^{-2}$ (Fig. 3**c**), $m^* = 0.23 \pm 0.01\ m_e$ is obtained from Lorentzian fits to the MDC peaks (Fig. 3**e**). Here, since the band bottom peak position is not clearly resolved, we estimated the effective mass by fitting the MDC peak positions over an energy window extending several hundred meV below $E_F$ with a quadratic function. Interestingly, this effective mass is lighter than that observed at high $n_{2D}$ (Figs. 1**d**, 3**d**, and 3**f**), and is close to the calculated band mass of $m_{DFT} = 0.26\ m_e$. It should be noted that single-shot $G_0W_0$ calculations, which include quasiparticle self-energy corrections beyond conventional DFT, improve the band gap but leave the conduction-band dispersion near the CBM nearly unchanged (Fig. 3**f**) (see also Fig. S11 of Supplementary Information), and thus cannot account for the band structure observed at high $n_{2D}$. We note that unresolved incoherent spectral weight, including possible phonon shake-off replicas, can make the apparent band bottom deeper and thereby lead to an underestimation of the effective mass in such wide-range fitting. To examine this uncertainty, we additionally evaluated the effective mass using a near-$E_F$ Fermi-velocity analysis and two-dimensional phenomenological spectral-function fitting with different fitting windows (see Supplementary Table S1 and Supplementary Figs. S2 and S3). These analyses show that the absolute mass obtained from wide-range fitting is sensitive to the high-binding-energy tail, but that the increase of $m^*$ with $n_{2D}$ remains robust irrespective of the evaluation method. The increase of $m^*$ with $n_{2D}$ (Fig. 3**g**) contrasts sharply with earlier findings for oxide wide-bandgap semiconductors (anatase $TiO_2$ and $SrTiO_3$ surfaces), where $m^*$ decreases toward the bulk band mass with increasing carrier density[7–9]. In those materials, electron–phonon coupling has been identified as the primary origin of the mass enhancement; this interaction—and the accompanying mass increase—diminishes at high carrier densities because of stronger electronic screening. Our ARPES results indicate that the conventional picture based solely on density-dependent screening of electron–phonon coupling cannot explain the mass renormalization at high $n_{2D}$ in β-$Ga_2O_3$.

The ARPES result at high $n_{2D}$ reveals band-renormalization signatures extending well beyond 100 meV in binding energy (Fig. 3**f**). Because the dominant LO-phonon energies in β-$Ga_2O_3$ lie below approximately 100 meV (Refs. 31,33), the renormalization caused by coupling to these phonons is expected within this characteristic energy scale ($0 < E_B < 100$ meV). The persistence of the band renormalization to higher binding energies ($E_B \sim 0.34$ eV), therefore, suggests that LO-phonon coupling alone is insufficient to account for the observed behavior. Peelaers *et al.* predicted an increase in the electron effective mass at high carrier density due to band non-parabolicity[34]. However, this model does not align with our ARPES observations, which show that a simple parabolic dispersion well reproduces the MDC peaks over the broad energy range $0 \leq E_B \leq 0.34$ eV (Fig. 3**f**). Although the absolute band-bottom position can be affected by the high-binding-energy spectral tail, the MDC peak dispersion near $E_F$ and the restricted-range spectral-function analysis consistently support the enhanced effective mass at high $n_{2D}$ (see Supplementary Table S1). Even when considering band non-parabolicity, the calculated effective mass of 0.34-0.37 $m_e$ at $k_F$ (= 0.25 Å$^{-1}$) (see Fig. S12 of Supplementary Information) is insufficient to explain the larger renormalization.

## Discussion

As discussed above, the increase of $m^*$ with $n_{2D}$ cannot be explained by the conventional screening-based picture commonly applied to oxide-semiconductor 2DESs. In that picture, increasing carrier density screens electron–phonon interactions and reduces the associated mass enhancement, in contrast to the trend observed here (Fig. 3). Thus, other mechanisms should be considered.

One might expect that electrons are partially trapped underneath the adsorbed $Cs^+$ ions, and the in-plane carrier trapping would reduce transport properties. However, because the $Cs^+$ ions are randomly adsorbed on the surface, such carrier trapping should cause significant broadening of 2DES spectral peaks; this contradicts the presence of a well-resolved parabolic band, indicative of the coherent propagation of electrons in the periodic potential. A possible interpretation is confinement-enhanced many-body interactions. In correlated or metallic oxide films, $m^*$ is known to increase with decreasing film thickness[36]. Previous ARPES studies of quasi-2D electron gases on InAs, CdO, and InN surfaces have also shown that many-body self-energy effects can substantially renormalize the energies of quantized conduction subbands[37], while RPA calculations for quasi-2D electron gases have shown that finite layer thickness suppresses the Coulomb-interaction-induced quasiparticle mass renormalization relative to the strictly 2D limit[38]. In the present β-$Ga_2O_3$ 2DES, increasing $n_{2D}$ shifts the charge centroid toward the surface and narrows the confinement width in our self-consistent model (see Supplementary Fig. S13). The resulting increase in the carrier concentration within the confined layer (estimated carrier-confinement depth of approximately 1.2 nm; Figs. 2**c**-2**e**) may enhance Coulomb interactions and modify collective modes such as plasmons. The associated electron–electron, electron–plasmon, or coupled phonon–plasmon self-energies may therefore contribute to the observed increase in $m^*$. Photoinduced excitonic correlations, including dark-exciton formation recently reported for an alkali-doped semiconductor under photoemission illumination[39], cannot be entirely excluded. In that study, the excitonic assignment was supported by a valence-band-derived spectral replica, an anisotropic conduction-band gap, and systematic doping- and temperature-dependent behavior. No corresponding excitonic signature is resolved in the present β-$Ga_2O_3$ spectra. Field-induced structural responses represent another possible contribution. Within our self-consistent Poisson–Schrödinger model, the estimated local interfacial field increases from 4.8 MV $cm^{-1}$ at $n_{2D}$ = $3.0 \times 10^{13}$ $cm^{-2}$ to 18 MV $cm^{-1}$ at high $n_{2D}$. Because these values are derived within the model and are sensitive to the adopted effective mass $m^*$, whose extracted value depends on the fitting method (Supplementary Table S1), they should be regarded as model-dependent estimates rather than directly determined quantities. Although the high-density estimate is numerically larger than the commonly quoted theoretical bulk breakdown field of β-$Ga_2O_3$ (approximately 8 MV $cm^{-1}$; Ref. 18), the two values are not directly comparable: the former is a local surface-field estimate within a confinement model, whereas the latter is a bulk breakdown criterion. Such a strong near-surface field could, in principle, modify the local lattice relaxation and in-plane bandwidth; however, the present measurements provide no direct structural evidence for this possibility. The present ARPES data do not permit a unique identification of its microscopic origin. Therefore, further targeted experimental and theoretical studies are required to determine the origin of the carrier-density-dependent increase in the ARPES-derived effective mass.

The direct observation of a 2DES at the surface of β-$Ga_2O_3$ provides a new perspective on the electronic properties of UWBG semiconductors. Notably, the formation of a 2DES is not a universal feature among semiconductors; it depends sensitively on the specific material and its crystal structure. For example, a 2DES readily appears on anatase $TiO_2$ (Refs. 9,40), yet has not been observed for the more stable rutile polymorph, where carriers tend to localize as small polarons[41]. Our ARPES results directly demonstrate that a metallic 2DES can form at the β-$Ga_2O_3$ surface despite its 4.7-eV band gap.

The ARPES data show the isotropic nature of this 2DES. This is a crucial insight, as it implies that the strong anisotropies in electrical conductivity previously reported for β-$Ga_2O_3$ (Ref. 21) do not originate solely from the intrinsic band structure of the mobile electrons. Instead, these transport anisotropies likely arise from extrinsic factors, such as anisotropic scattering and crystallinity[42]. Furthermore, the highest $n_{2D}$ of $1.0 \times 10^{14}$ $cm^{-2}$ is an order of magnitude higher than previously reported for β-$Ga_2O_3$ interfaces[43–45]. Although Cs deposition on a cleaved surface under UHV is not a practical device-doping strategy, it provides experimental access to the high-density 2DES regime, enabling fundamental studies of density-dependent electronic properties that may inform future research on high-power and high-frequency β-$Ga_2O_3$ devices[15]. The dense 2DES is also expected to push the plasmon frequency firmly into the terahertz (THz) domain, which could motivate further exploration of plasmon-related phenomena, including Dyakonov-Shur-type physics [46,47].

Finally, our findings establish β-$Ga_2O_3$ as a promising new platform for quantum materials research. The low-temperature electron mobilities exceeding 10,000 $cm^2$ $V^{-1}$ $s^{-1}$ reported for β-$Ga_2O_3$ (Ref. 48) further highlight its potential for quantum transport studies. In this context, the combination of a surface 2DES, expected strong carrier confinement, and inversion-symmetry breaking motivates future investigations of possible spin splitting[6] and collective phenomena. These results thus extend the exploration of quantum phenomena into the UWBG-semiconductor regime.

## Conclusion

In summary, micro-focused ARPES directly resolved the momentum-space electronic structure of a 2DES on the UWBG semiconductor β-$Ga_2O_3$. With alkali-metal deposition and synchrotron-radiation-based ARPES measurements on high-quality β-$Ga_2O_3$ single crystals, we directly resolved a well-defined dispersive 2DES band. Our ARPES results further show that the 2DES is essentially isotropic in-plane, suggesting that the reported *b*- and *c*-axis transport anisotropy does not arise solely from its band dispersion. Our approach further enables tuning of $n_{2D}$ up to $1.0 \times 10^{14}$ $cm^{-2}$, substantially extending the accessible density range for 2DESs on β-$Ga_2O_3$. Most notably, the $m^*$ of the 2DES increases by nearly a factor of two with increasing $n_{2D}$ in the high-$n_{2D}$ regime while the quasiparticle states remain coherent. This behavior is in marked contrast to the density-dependent mass reduction reported for 2DESs on wide-bandgap oxides such as $SrTiO_3$ and $TiO_2$. This establishes β-$Ga_2O_3$ as a new platform for investigating surface and interfacial quantum states in UWBG semiconductors.

## Acknowledgements

The authors are very grateful to Y. Kumagai for valuable discussions. The synchrotron radiation experiments were performed at BL06U of NanoTerasu with the approval of the Japan Synchrotron Radiation Research Institute (JASRI) (Proposal No. 2025A9038). ARPES measurements at KEK-PF were approved by the Program Advisory Committee (Proposal No. 2023G083) at the Institute of Materials Structure Science, KEK. Part of this work was performed under the Use-of-UVSOR Synchrotron Facility Program (Proposal No. 24IMS6836) of the Institute for Molecular Science, National Institutes of Natural Sciences. Some of the computational results in this study were obtained using supercomputing resources at the Cyberscience Center, Tohoku University.

## Funding

This work was supported by JSPS Grants-in-Aid for Scientific Research (Grant Nos. 23K25803 and 23K17878), the MEXT Program for Creation of Innovative Core Technology for Power Electronics (Grant No. JPJ009777), and a JSPS Grant-in-Aid for Scientific Research (A) (Grant No. 25H00725).

## Author contributions

ARPES measurements were carried out by R.Y., S.S.P., YJL.L., Y.M., H.T., B.L., H.Y., T.A., J.Y., R.K., T.N., T.K., A.Yasui, M.Kitamura, H.I., K.H., and K.O. R.Y., S.S.P., and YJL.L. analyzed the data. H.K., T.K., A.Yasui, M.Kitamura, H.I., K.H., and K.O. were responsible for the synchrotron ARPES measurement setups used in the experiments. H.M. and A.K. performed the band structure calculations. R.Y. calculated the band bending structures. V.V.K., T.T., M.Kitahara, T.H., and A.Yoshikawa grew the crystals and determined the optical band gap. R.Y. and N.S. were responsible for overall project planning and direction. R.Y. wrote the manuscript with contributions by H.M., S.S.P., YJL.L., Y.M., N.S., N.C., T.S., M.C-R., Y.K., H.K., and K.O. All authors contributed to the scientific discussion of the results.

## Competing interests

The authors declare no competing interests.

## Methods

### Crystal growth

The β-$Ga_2O_3$ single crystals used in this study were grown by the recently developed OCCC method[24]. β-$Ga_2O_3$ (99.99% purity) powder was used as the starting material. An air atmosphere was used for the crystal growth at a flow rate in the range of 5–10 L/min. The dimensions of the water-cooled Cu basket were 85 mm × 60 mm. Yttria-stabilized zirconia was used for the top side insulation. The raw material, a few centimeters thick near the water-cooled Cu basket, was sintered, while the rest became molten. The melt was held by the sintered raw material. A seed crystal was touched to the center of the melt from above, and after it stabilized, a single crystal was grown by pulling it up while rotating it. The crystal growth rate was set to 3–5 mm/h, and the rotation speed was set to 4–6 rpm. β-$Ga_2O_3$ crystals, grown previously by means of the floating-zone method along the ⟨010⟩ axis, were used as the seeds. The obtained β-$Ga_2O_3$ single crystal had a diameter ≤ 46 mm. Impurities were determined using glow discharge mass spectrometry (GDMS) analysis. Chemical analysis detected only raw-material impurities, with Cu, Zr, and noble-metal contaminants below instrument limits. As expected, there were no external impurities, since the raw powder is sintered to become the container that holds the melt. X-ray rocking-curve widths were comparable to those of edge-defined film-fed-growth crystals, confirming high structural quality.

## ARPES measurements

ARPES data were acquired mainly at beamline BL06U of NanoTerasu, where a synchrotron radiation beam with a spot size of $20 \times 10\ \mu m^2$ was used; data shown in Fig. 1**f** were obtained at beamline BL-2A of the KEK Photon Factory, where a larger beam spot size ($500 \times 200\ \mu m^2$) was used. The β-$Ga_2O_3$ single crystals were cleaved under ultra-high vacuum (UHV) along the *bc* plane[49] and dosed with Cs using a dispenser (SAES Getters) at room temperature. On oxide semiconductor surfaces, alkali-metal adsorption—owing to the large ionic radii—is unlikely to intercalate into the lattice, enabling stable electron donation from the surface[9]. After dosing, the samples were transferred to the ARPES measurement chambers. In ARPES measurements at both beamlines, the incident light and the analyzer slits were in the same plane, and *p*-polarized light was used for the excitation (see Supplementary Fig. S1 for the polarization-dependent ARPES intensity). The photon energies used were 92 eV for Fig. 1**e** and Figs. 3**a**-3**c**, which were taken at 2nd $\bar{\Gamma}$, and 102 eV for Fig. 1**d**, which was taken at the 1st $\bar{\Gamma}$, where the same dataset is shown in Fig. 3**d**. At these photon energies, relatively strong ARPES intensities from the 2DES were obtained. The samples were set so that the *x* axis (*b* axis) was parallel to the analyzer slit and cooled to ≤ 20 K during the measurements. Because the crystals were undoped, slight surface charging was present; the band edges of the metallic states were therefore referenced to $E_F$ (see Supplementary Figs. S5 and S7 for the temporal stability and the effective energy broadening of the ARPES spectra and the reproducibility of the ARPES-derived effective masses under different experimental conditions). A work function of 4.11 eV (Ref. 50) and an inner potential of 15 eV (Ref. 51) were assumed when converting the ARPES spectra into momentum space. ARPES measurements were also conducted at BL5U of UVSOR-III to optimize electron-doping conditions. Sheet carrier densities are derived from $\pi k_F^2/(2\pi^2)$ by assuming circular Fermi surfaces.

## DFT calculations

We performed DFT calculations[52], as implemented in the Vienna ab initio simulation package (VASP)[53,54]. The Perdew–Burke–Ernzerhof (PBE) exchange-correlation functional[55] was chosen for electronic structure calculations and geometry relaxation. The electron–ion interaction was described by the projector-augmented-wave (PAW) method[56,57]. The Brillouin zone was sampled using an automatically generated Monkhorst–Pack mesh, with the $k$-point density chosen to ensure convergence of the total energy and structural optimization[58]. A plane-wave cutoff energy of 800 eV was used for the electronic wave functions. After structural optimization, the effective mass of the conduction band at the Γ point was determined in the $k_x$, $k_y$, and $k_z$ directions to align with the experimentally cleaved surface. Using the optimized structure, the PDOS and total density of states (DOS) were calculated. Furthermore, the band structure was plotted using wave vectors within the Brillouin zone, following the approach of the Van de Walle group[59]. The crystal structures in Figs. 1**a** and 2**e** were generated using VESTA[60].

### $G_0W_0$ calculations

In order to account for quasiparticle effects in the band structure calculations, we employed Hedin's *GW* approach[61] within the $G_0W_0$ scheme as implemented in VASP[62]. The electronic structure obtained from the PBE functional was used as the starting point for the $G_0W_0$ calculations. The crystal structure was optimized using a plane-wave cutoff energy of 820 eV, 196 bands, and an $11 \times 11 \times 11$ Monkhorst–Pack $k$-point mesh. In the subsequent $G_0W_0$ calculations, a plane-wave cutoff energy of 480 eV, 512 bands, and an $8 \times 8 \times 8$ Γ-centered Monkhorst–Pack $k$-point mesh were adopted, following previously reported values[63]. To obtain the band structure, we performed Wannier interpolation[64–67] of the results computed on the uniform $k$-point grid using the selected columns of the density matrix (SCDM) method[68] implemented in VASP, in combination with Wannier90 (Ref. 69). The number of Wannier orbitals was set to 120 to ensure convergence of the conduction band structure. The calculations were carried out using supercomputing resources at Cyberscience Center, Tohoku University.

### Self-consistent Poisson–Schrödinger calculations

Carrier-density profiles and confinement potentials (Figs. 2**c** and 2**d**) were obtained by self-consistently solving the Poisson–Schrödinger equations[70,71], using the isotropic effective mass determined from ARPES for all subbands, a bulk donor concentration of $2 \times 10^{14}$ cm$^{-3}$, and a relative permittivity of 12 for β-$Ga_2O_3$—the mid-range of reported bulk values[72]. The surface potential was iteratively adjusted until the calculated bottom of the first subband matched its experimental binding energy, after which the converged potential and charge density were recorded. Changing the assumed bulk donor concentration by two orders of magnitude had a negligible influence on the band-bending profile, carrier distribution, or subband energies.

## Data availability

Any data supporting the findings of this study are available from the corresponding author upon request.

## Figure captions

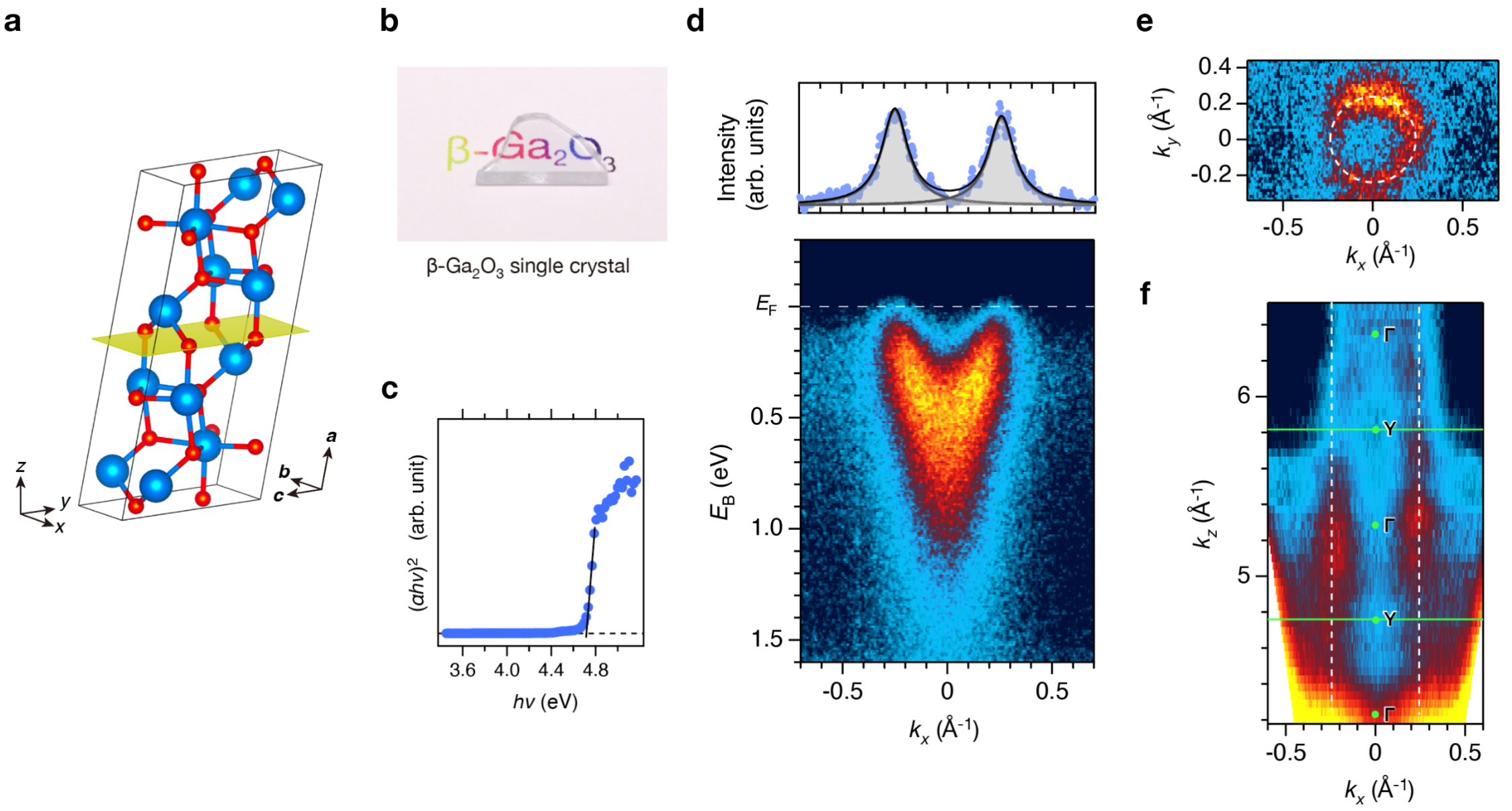

**Fig. 1 | Optical band gap and 2DES observed on β-$Ga_2O_3$. a**, Crystal structure of a $1 \times 1 \times 1$ conventional unit cell. The yellow plane indicates the ARPES measurement plane, parallel to both the *xy* and *bc* planes. The lattice parameters are $a$ = 12.23 Å, $b$ = 3.04 Å, and $c$ = 5.80 Å (Ref. 25). **b**, A photograph of a 1.7 mm-thick $\beta$-$Ga_2O_3$ single crystal, placed on printed paper. **c**, Energy dependence of $(\alpha h\nu)^2$. The optical band gap of the crystal is determined to be 4.7 eV from linear extrapolation of the fitted region of the slope. **d**, ARPES intensity map near the normal emission geometry for the electron-doped $\beta$-$Ga_2O_3$ surface. The momentum-distribution curve (MDC) at $E_F$ is plotted as blue dots above the ARPES map, and the black curve shows the fitting result obtained with two Lorentzian functions. **e**,**f**, constant-energy ARPES intensity maps at the Fermi level ($E_F$) taken for the $k_x$-$k_y$ plane (**e**) and the $k_x$-$k_z$ plane (**f**). Here, the intensities are integrated over the energy window of $\pm 20$ meV. The horizontal green lines mark the approximate bulk Brillouin-zone boundary (see Supplementary Fig. S10) at $k_y$ = 0 Å$^{-1}$, and the high-symmetry Γ and Y points are labelled in the figure.

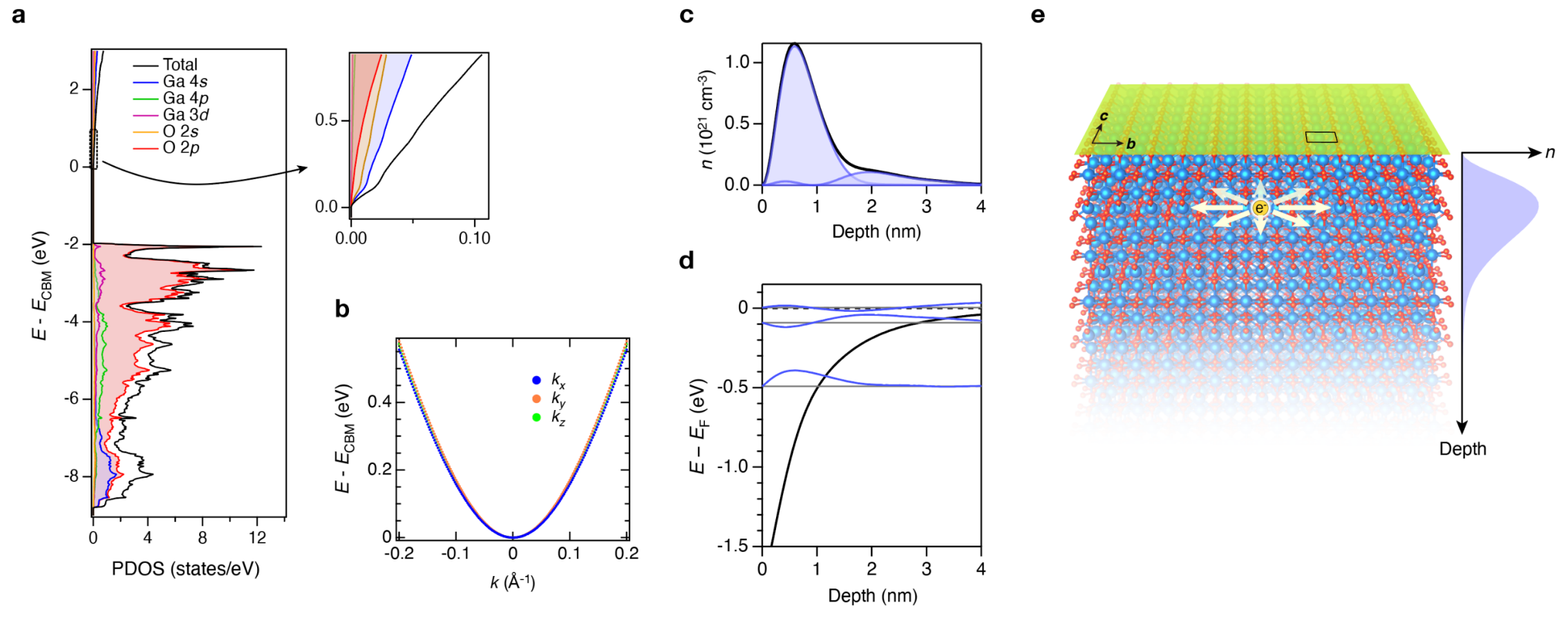

**Fig. 2 | Calculated electronic structure and confinement of β-$Ga_2O_3$. a**, PDOS of bulk *β*-$Ga_2O_3$, where 0 eV is set to the energy of the conduction band minimum ($E_{CBM}$). Expanded plots near the CBM are shown on the right side. **b**, Calculated band structures near the CBM along the $k_x$, $k_y$, and $k_z$ directions, where 0 $Å^{-1}$ corresponds to the Γ point of the Brillouin zone. **c**,**d**, Calculation results for the density profiles (**c**) and the potential curve (**d**). The black and blue curves in **c** correspond to the total and partial electron densities, respectively. The gray lines in **d** represent the subband energy minima, where the corresponding eigenfunctions are also shown as blue curves. **e**, Schematic illustration of electron transport for the observed two-dimensional metallic states. Yellow arrows represent the isotropic conduction of the 2DES in the *bc* plane (*xy* plane).

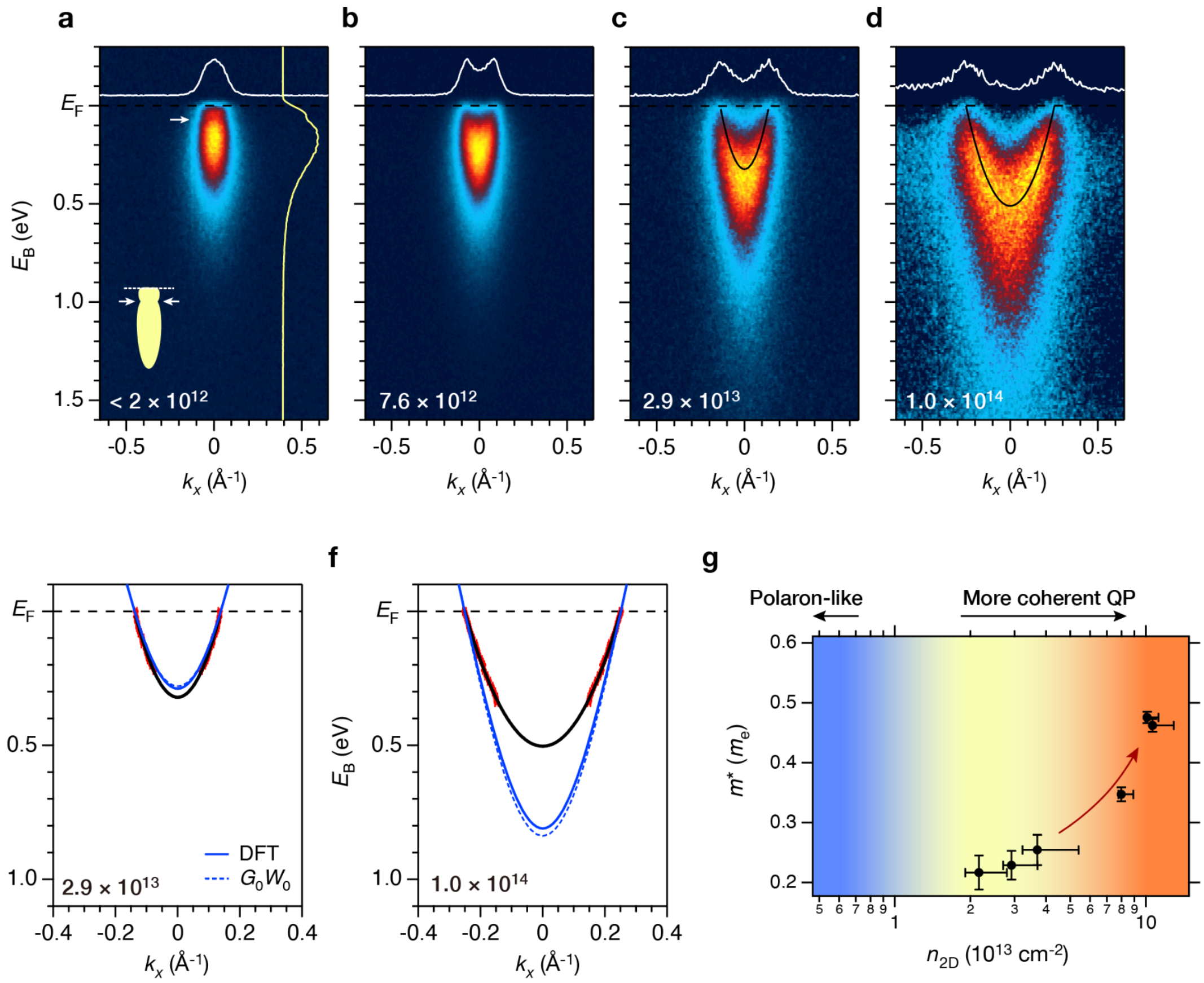

**Fig. 3 | Evolution of the 2DES against electron doping. a-d**, ARPES intensity plots measured at different sheet carrier densities, $n_{2D}$. MDCs integrated over a 20-meV energy window centered at $E_F$ are shown above the corresponding plots. The yellow curve in **a** is the energy distribution curve taken at $k_x$ = 0 Å$^{-1}$ integrated in a range of $\pm 0.04$ Å$^{-1}$. A schematic outline of the ARPES spectral contour, together with the position of the dip feature, is superimposed in **a**. **e,f,** MDC peak positions obtained by fitting the positive- and negative-$k_x$ peaks at each energy in **c** and **d**, respectively (red circles). The error bars indicate the corresponding fitting uncertainties. Each panel represents $n$ = 1 independent ARPES experiment. The blue curves represent the calculated band structure, whose energy is shifted so that the band crosses $E_F$ at the experimentally obtained $k_F$. Parabolic curves obtained from curve fitting are overlaid as black curves in **c-f**. The $n_{2D}$ values, given in units of cm$^{-2}$, were calculated from the MDC-derived $k_F$ values and are indicated in the lower-left corner of each panel in **a–f**. **g**, Plot of the obtained $m^*$ against $n_{2D}$. Each data point represents $n$ = 1 independent ARPES

experiment. The plotted $n_{2D}$ values were calculated from the MDC-derived $k_F$, whereas the horizontal error bars incorporate both the MDC fitting uncertainty and the method-dependent variation between the MDC- and gradient-derived[35] $k_F$ values, including the variation obtained with different initial fitting parameters. The vertical error bars span the range of effective masses obtained from repeated parabolic fits to dispersions determined from energy-dependent MDC peak positions. Both error bars describe the spread of estimates obtained from the same underlying dataset, rather than standard deviations, standard errors, or confidence intervals. Repeated fits to the same dataset are not independent experimental replicates. The background color and upper labels schematically indicate a gradual evolution from polaron-like spectral features at low $n_{2D}$ toward a more coherent quasiparticle (QP) response at high $n_{2D}$.

# Supplementary Information

# Anomalous mass renormalization in two-dimensional electron systems on ultra-wide bandgap semiconductor β-$Ga_2O_3$

**Ryu Yukawa,[1,*] Hiroshi Mizuseki,[2] Suryo Santoso Putro,[3] Yé-Jin L. Lee,[1,4] Yuuki Masutake,[3] Hinako Telengut,[3] Boxuan Li,[3] Hajime Yamamoto,[3] Tadashi Abukawa,[1,3] Junya Yoshida,[1] Vladimir V. Kochurikhin,[5,6] Taketoshi Tomida,[5,6,7] Masanori Kitahara,[5,6,7] Takahiko Horiai,[6,8] Akira Yoshikawa,[5,6,7,8] Nobuhiko Sarukura,[8,9] Noriko Chikumoto,[9] Toshihiko Shimizu,[8,9] Marilou Cadatal-Raduban,[9,10] Yoshiyuki Kawazoe,[8,11,12] Ryuhei Kohno,[1] Hiroshi Kumigashira,[3] Takuto Nakamura,[13,14] Tatsuhiko Kanda,[15] Akira Yasui,[15] Akihiro Koide,[15] Miho Kitamura,[16] Hideaki Iwasawa,[16,17,18] Koji Horiba,[16] Kenichi Ozawa[19]**

[1]International Center for Synchrotron Radiation Innovation Smart (SRIS), Tohoku University, Sendai 980-8572, Japan

[2]Korea Institute of Science and Technology (KIST), Seoul 02792, Republic of Korea

[3]Institute of Multidisciplinary Research for Advanced Materials (IMRAM), Tohoku University, Sendai 980-8577, Japan

[4]Faculté des Sciences et Ingénierie, Sorbonne Université, F-75005 Paris, France

[5]Institute for Materials Research (IMR), Tohoku University, Sendai 980-8577, Japan

[6]C&A Corporation, Sendai 980-0811, Japan

[7]FOX Corporation, Sendai 982-0024, Japan

[8]New Industry Creation Hatchery Center (NICHe), Tohoku University, Sendai 980-8579, Japan

[9]Institute of Laser Engineering, Osaka University, 2-6 Yamadaoka, Suita 565-0871, Japan

[10]Unitec Institute of Technology, 139 Carrington Road, Mount Albert, Auckland 1025, New Zealand

[11]Department of Physics, SRM University-AP, Amaravati 522 240, Andhra Pradesh, India

[12]School of Physics, Institute of Science and Center of Excellence in Advanced Functional Materials, Suranaree University of Technology, Nakhon Ratchasima 30000, Thailand
[13]Graduate School of Frontier Biosciences, Osaka University, Suita 565-0871, Japan
[14]Department of Physics, Graduate School of Science, Osaka University, Toyonaka 560-0043, Japan
[15]Japan Synchrotron Radiation Research Institute (JASRI), Sayo 679-5198, Japan
[16]NanoTerasu Center, National Institutes for Quantum Science and Technology (QST), Sendai 980-8572, Japan
[17]Synchrotron Radiation Research Center, National Institutes for Quantum Science and Technology (QST), Sayo 679-5148, Japan
[18]Research Institute for Synchrotron Radiation Science, Hiroshima University, Higashi-Hiroshima 739-0046, Japan
[19]Institute of Materials Structure Science, High Energy Accelerator Research Organization (KEK), Tsukuba 305-0801, Japan

* e-mail: r.yukawa@tohoku.ac.jp

**Polarization dependence of the ARPES intensity**

Figure S1 shows the polarization dependence of the ARPES intensity from the 2DES formed on the electron-doped β-$Ga_2O_3$ surface. The measurements were performed at Photon Factory BL-2A using the experimental geometry schematically illustrated in Fig. S1**a**. The ARPES intensity map taken with linear horizontal (LH; *p*) polarization shows a clear 2DES band near $E_F$ (Fig. S1**b**), whereas the corresponding signal is strongly suppressed with linear vertical (LV; *s*) polarization (Fig. S1**c**). These spectra were recorded at the same position on the sample surface before and after switching the polarization, confirming that the observed contrast originates from the polarization dependence rather than from spatial inhomogeneity of the doped surface. This pronounced polarization dependence is attributed to a photoemission dipole-matrix-element effect[1]. In the present geometry, LH and LV correspond to *p*- and *s*-polarized light, respectively, and couple differently to the predominantly *s*-like conduction-band states. The strong suppression under LV polarization should therefore be regarded as a geometry-dependent matrix-element effect.

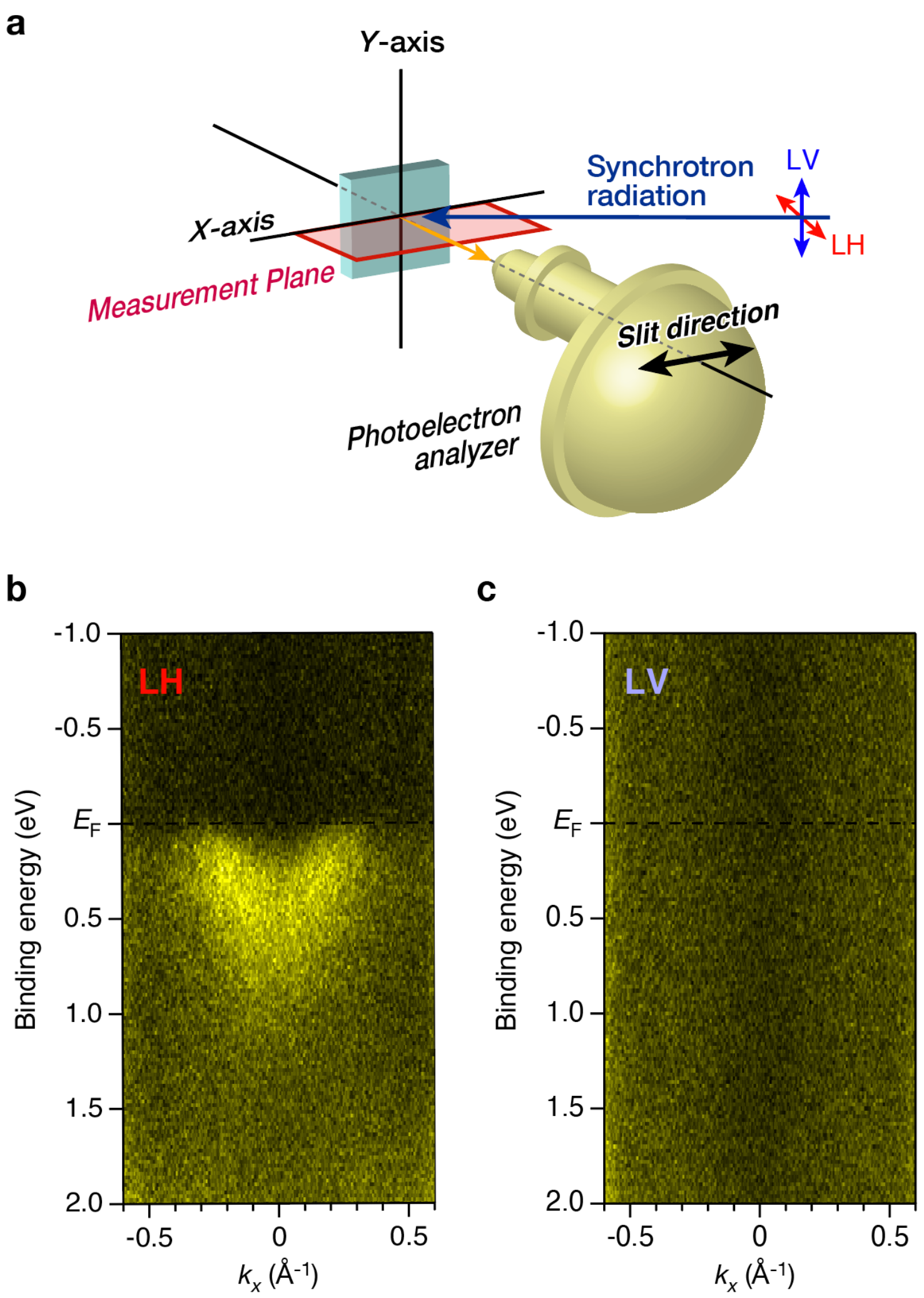

**Fig. S1 I Experimental geometry and polarization-dependent ARPES intensity of the β-$Ga_2O_3$ surface 2DES. a**, Schematic of the experimental geometry for the polarization-dependent ARPES measurements performed at Photon Factory BL-2A. The red plane indicates the measurement plane. LH and LV denote linear horizontal and linear vertical polarizations, respectively. **b,c**, ARPES intensity maps measured near the normal-emission geometry with LH (**b**) and LV (**c**) polarizations at a photon energy of 92 eV. A clear 2DES band is observed with LH polarization, whereas the corresponding intensity is strongly suppressed with LV polarization. The spectra in **b** and **c** were acquired at the same position on the sample surface before and after switching the undulator polarization.

**Effective-mass analysis**

To examine the robustness of the carrier-density-dependent effective-mass enhancement, we evaluated the effective mass using three complementary approaches for the low-, intermediate-, and high- $n_{2\mathrm{D}}$ spectra, corresponding to $n_{2\mathrm{D}} = 2.9 \times 10^{13}$ , $8.0 \times 10^{13}$ , and $10.1 \times 10^{13}\ \mathrm{cm}^{-2}$, respectively. For comparison with the two-dimensional spectral-function model, the ARPES intensity maps were symmetrized with respect to $k_x = 0$ Å$^{-1}$, thereby reducing the intensity asymmetry mainly caused by photoemission matrix-element effects.

**1. Parabolic fitting of the MDC peak positions**

Momentum-distribution curves were fitted with Lorentzian functions over a relatively wide energy range where the dispersive MDC peak could be reliably identified. The resulting peak positions were fitted using a parabolic dispersion:

$$E(k) = E_0 + \frac{\hbar^2 (k - k_0)^2}{2m^*}.$$

Here, $E_0$, $k_0$, and $m^*$ denote the band-bottom energy, band-center position, and effective mass, respectively. This fitting procedure yielded effective masses ranging from $0.23\ m_e$ to $0.48\ m_e$, corresponding to the values reported in the original manuscript. In this analysis, we did not use the broad high-binding-energy tail itself as a peak position; instead, the parabolic fit was performed using the MDC peak positions in the dispersive region, thereby largely reducing the influence of unresolved incoherent spectral weight and the ambiguous band-bottom region.

**2. Effective mass derived from the near-$E_F$ Fermi velocity**

To reduce the influence of the poorly resolved band-bottom region, the effective mass was independently evaluated using only the MDC peak positions near $E_F$. The Fermi velocity $v_F$ was obtained from a linear fit to the near-$E_F$ dispersion, and the Fermi-surface effective mass was calculated as

$$m_{\mathrm{F}}^* = \frac{\hbar k_{\mathrm{F}}}{v_{\mathrm{F}}}.$$

To enable a direct comparison among the three carrier-density conditions, the Fermi velocity was determined by linearly fitting the MDC peak positions over the same binding-energy range of 0 $\leq E_B \leq 0.10$ eV for all spectra. The resulting effective masses were $0.20 \pm 0.07\ m_e$, $0.28 \pm 0.06\ m_e$, and $0.50 \pm 0.09\ m_e$ for the low-, intermediate-, and high-$n_{2D}$ spectra, respectively.

**3. Two-dimensional spectral-function fitting**

The momentum-symmetrized ARPES intensity maps were analyzed by fitting the full $k_x$–energy intensity distribution using a phenomenological spectral-function model based on established ARPES spectral-function analyses and two-dimensional fitting approaches[2–4]. Defining $\omega = E - E_F$, the spectral function was represented by a single Lorentzian component,

$$A(k_x, \omega) = \frac{1}{\pi} \frac{\Gamma(\omega, T)}{[\omega - \xi(k_x)]^2 + \Gamma(\omega, T)^2},$$

centered on the parabolic dispersion,

$$\xi(k_x) = \frac{\hbar^2}{2m^*}(k_x^2 - k_F^2).$$

This definition places the Fermi crossings at $k_x = \pm k_F$. The energy-dependent Lorentzian half-width at half maximum was modeled phenomenologically as

$$\Gamma(\omega, T) = \Gamma_0 + \alpha[\min(|\omega|, \omega_c)^2 + (\pi k_B T)^2],$$

where $\Gamma_0$ is the energy-independent residual Lorentzian half-width at half maximum, $\alpha$ is the coefficient governing the quadratic energy and temperature dependence of the linewidth, and $\omega_c$ is the cutoff energy above which the energy-dependent contribution is held constant. The quadratic dependence provides a phenomenological Fermi-liquid-like description of the low-energy linewidth. Above $\omega_c$, the energy-dependent contribution is held constant to avoid extrapolating the quadratic form into the high-binding-energy region. The cutoff was located above the upper binding-energy limits of the restricted fitting windows and therefore does not affect the restricted-range results. A single common set of linewidth parameters was used for all three carrier-density conditions to avoid introducing density-dependent changes in the assumed spectral broadening. These parameters were used only to describe the overall line shape and were not interpreted as a quantitative microscopic self-energy. No separate incoherent spectral component was included to provide a consistent analysis. The calculated photoemission intensity was expressed as

$$I_{calc}(k_x, \omega) = C[A(k_x, \omega) f(\omega, T)] \otimes G_E \otimes G_k + b_0 + b_1\omega,$$

where $f(\omega, T)$ is the Fermi–Dirac distribution, $G_E$ and $G_k$ are Gaussian energy and momentum broadening functions, respectively, $C$ is an overall intensity scale, and $b_0 + b_1\omega$ is a momentum-independent background linear in energy. The temperature and Gaussian energy-broadening FWHM were fixed according to the experimental conditions and total experimental energy resolution, respectively. For the fits shown in Fig. S2, $k_F$ was fixed at 0.135, 0.224, and 0.252 $Å^{-1}$ for the low-, intermediate-, and high-$n_{2D}$ spectra, respectively, using values determined from the experimental MDC peaks near $E_F$. The effective mass $m^*$ and effective Gaussian momentum broadening were varied. For the additional restricted-range fits shown in Fig. S4, $k_F$ was varied together with these parameters. For each parameter set, a calculated $k_x$–energy intensity map was compared with the experimental map. This full-map fitting was used as a phenomenological forward-modeling analysis to test the robustness of the carrier-density-dependent effective-mass trend, rather than to quantitatively determine the microscopic self-energy. Two energy windows were examined. The wide-range fits extended from 70 meV above $E_F$ to binding energies of 1.0 eV for the low- $n_{2D}$ spectrum and 1.3 eV for the intermediate- and high-$n_{2D}$ spectra (Fig. S2**c**). To reduce the influence of the broad high-binding-energy intensity near and below the apparent band-bottom region, restricted-range fits were performed down to binding energies of 0.20, 0.25, and 0.30 eV for the low-, intermediate-, and high-$n_{2D}$ spectra, respectively (Fig. S2**b**). All fits were performed over $-0.65 \leq k_x \leq 0.65$ Å$^{-1}$. The $k_F$ varied fits used the same restricted energy windows and yielded $k_F$ = 0.139, 0.240, and 0.258 $Å^{-1}$ and $m^*$ = 0.24, 0.48, and 0.55 $m_e$ for the low-, intermediate-, and high-$n_{2D}$ spectra, respectively (Fig. S4 and Table S1).

At first glance, the wide-range fits for the low-$n_{2D}$ spectrum in Fig. S2**c** appear to reproduce the overall intensity distribution over the displayed energy range more closely than the restricted-range fits in Fig. S2**b**. However, the wide-range fit places the calculated MDC peaks at slightly larger $|k_x|$ than the experimental peaks (Fig. S3**b**). The corresponding EDC comparison in Fig. S3**c** shows that the experimental spectrum contains broad intensity extending to higher binding energy near the apparent band-bottom region. Because the present single-component model does not independently represent this broad tail, the wide-range fit attempts to accommodate it by shifting the fitted band bottom to higher binding energy. For a parabolic band with fixed $k_F$, the band-bottom binding energy is

$$E_{\mathrm{B,bottom}} = \frac{\hbar^2 k_{\mathrm{F}}^2}{2m^*}.$$

A fitted band bottom at higher binding energy therefore produces a smaller apparent effective mass and shifts the calculated MDC peaks outward in momentum. This effect is not limited to the

low-$n_{2D}$ spectrum: including the band-bottom and high-binding-energy regions also yields systematically smaller masses for the intermediate- and high-$n_{2D}$ spectra. The wide-range fits yielded best-fit masses of 0.15 $m_e$, 0.29 $m_e$, and 0.37 $m_e$, whereas the corresponding restricted-range fits yielded 0.22 $m_e$, 0.34 $m_e$, and 0.50 $m_e$ for the low-, intermediate-, and high-$n_{2D}$ spectra, respectively. Thus, fitting the broad intensity near the apparent band bottom biases the fitted dispersion toward a lighter effective mass. The present spectra do not establish the microscopic origin of the broad high-binding-energy tail. The two-dimensional fits were therefore used to assess the mass trend and obtain model-dependent estimates of $k_F$, rather than to identify the origin of this spectral weight.

To minimize the influence of the ambiguous band-bottom region and reduce model dependence, the effective masses presented in Fig. 3**g** of the main text were obtained by fitting the experimentally extracted MDC peak positions in the well-defined dispersive region closer to $E_F$ with a simple parabolic dispersion. In Fig. S4**b**, the calculated near-$E_F$ MDC peaks approximately coincide with the experimental peaks for the low- and high-$n_{2D}$ spectra but lie at slightly larger $|k_x|$ for the intermediate-$n_{2D}$ spectrum. The intermediate fitted $k_F$ and $m^*$ also exceed the MDC-based values of 0.224 Å$^{-1}$ and 0.35 $m_e$, illustrating the model dependence of the absolute parameters. As summarized in Table S1, all five analyses—the two MDC-based analyses (a parabolic fit to the peak positions and a near-$E_F$ Fermi-velocity analysis) and the three two-dimensional spectral-function fits (wide-range with $k_F$ fixed, restricted-range with $k_F$ fixed, and restricted-range with $k_F$ varied)—yield a monotonic increase in the best-fit effective mass with increasing $n_{2D}$. Although the absolute mass depends on the fitting procedure and energy window, the carrier-density-dependent increase in $m^*$ is consistently reproduced.

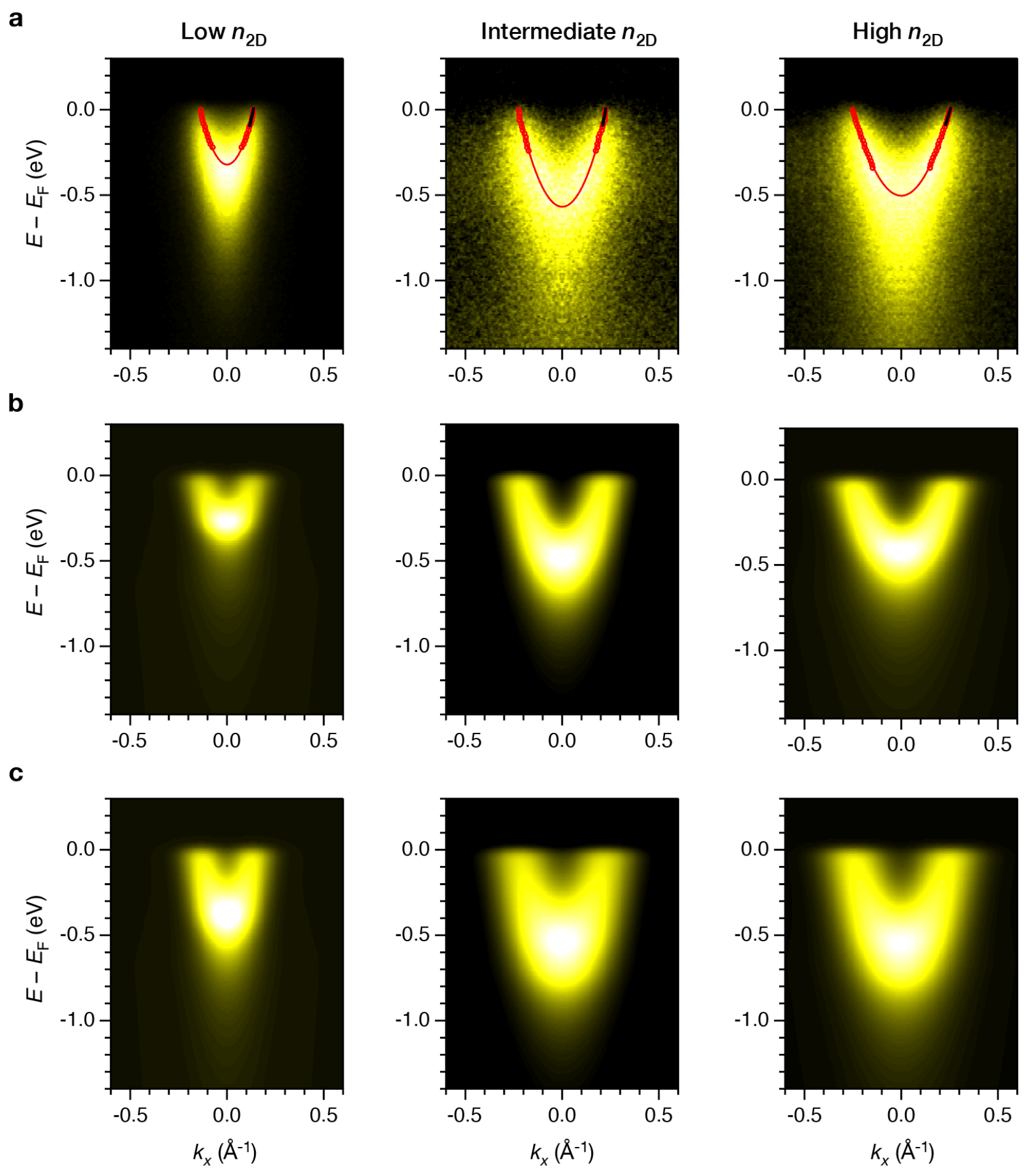

**Fig. S2 I Effective-mass analysis using MDC peak positions and two-dimensional spectral-function fitting. a**, Symmetrized ARPES intensity maps for the low-, intermediate-, and high-$n_{2D}$ spectra. The maps were symmetrized with respect to $k_x = 0$ Å$^{-1}$. Red open circles indicate the MDC peak positions obtained by Lorentzian fitting. The red curves represent parabolic fits to the MDC peak positions over a wide energy range, whereas the black straight lines indicate linear fits to the near-$E_F$ MDC peak positions used to determine the Fermi velocity. **b**,**c**, Calculated ARPES intensity maps obtained from the two-dimensional spectral-function fits to the restricted (**b**) and wider (**c**) energy regions for the corresponding low-, intermediate-, and high-$n_{2D}$ spectra. The calculated maps include the Fermi–Dirac distribution, energy broadening, momentum broadening, an energy-dependent linewidth, and a linear energy-dependent background.

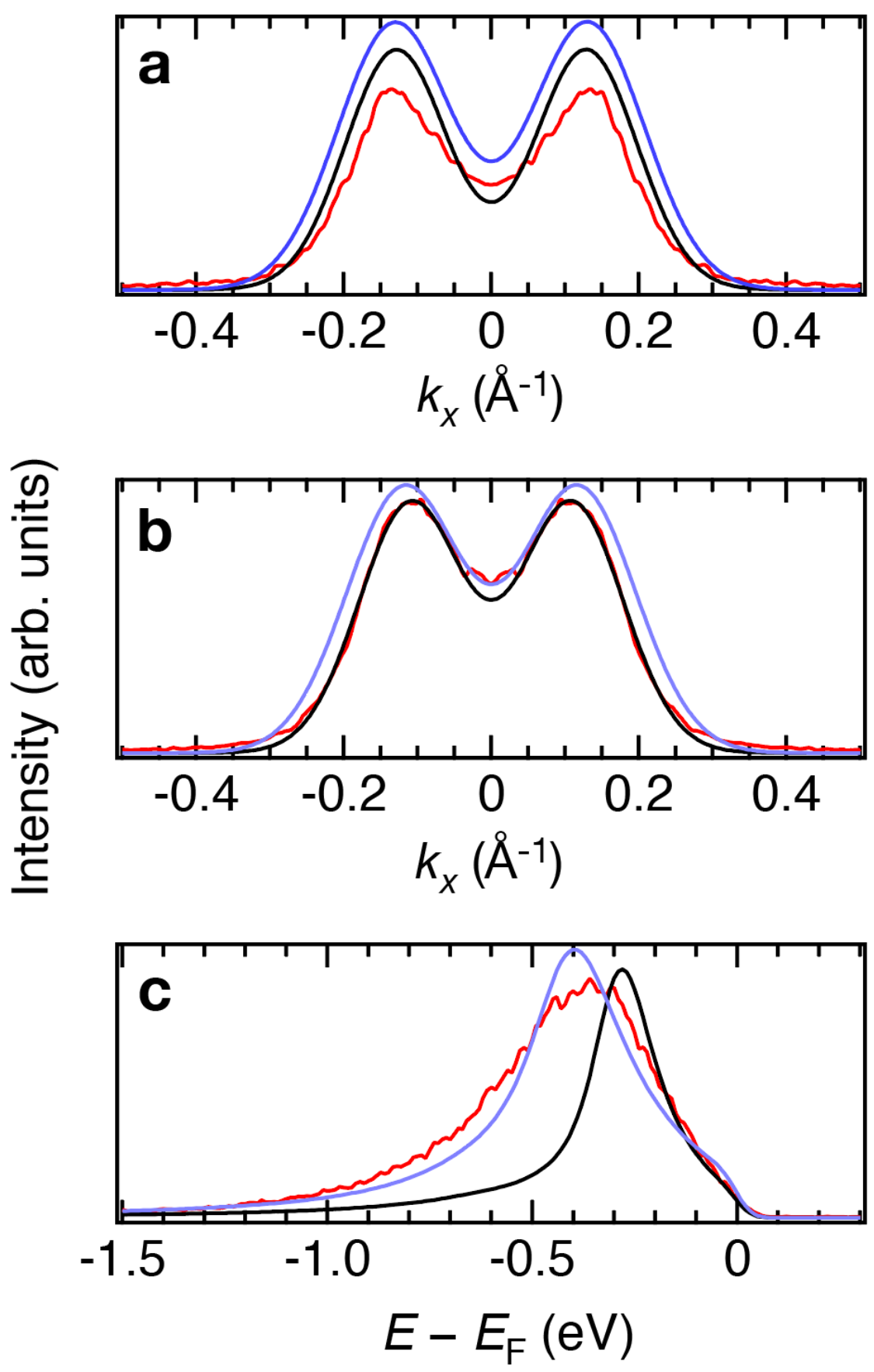

**Fig. S3 | Comparison between the experimental ARPES spectra and the two-dimensional spectral-function fit for the low-$n_{2D}$ spectrum.** Experimental spectra are shown by red curves, and the calculated spectra obtained from the two-dimensional spectral-function fit (Fig. S2**b**) are shown by black curves, whereas those obtained from the wide-energy-range fit (Fig. S2**c**) are shown by blue curves. **a**, MDCs at $E - E_\mathrm{F} = 0$ eV. **b**, MDCs at $E - E_\mathrm{F} = -0.10$ eV. For panels (**a**) and (**b**), the MDCs were obtained by integrating the intensity over an energy window of 20 meV. **c**, EDCs at $k_x = 0$ Å$^{-1}$, obtained by integrating the intensity over a momentum window of $0.02$ Å$^{-1}$.

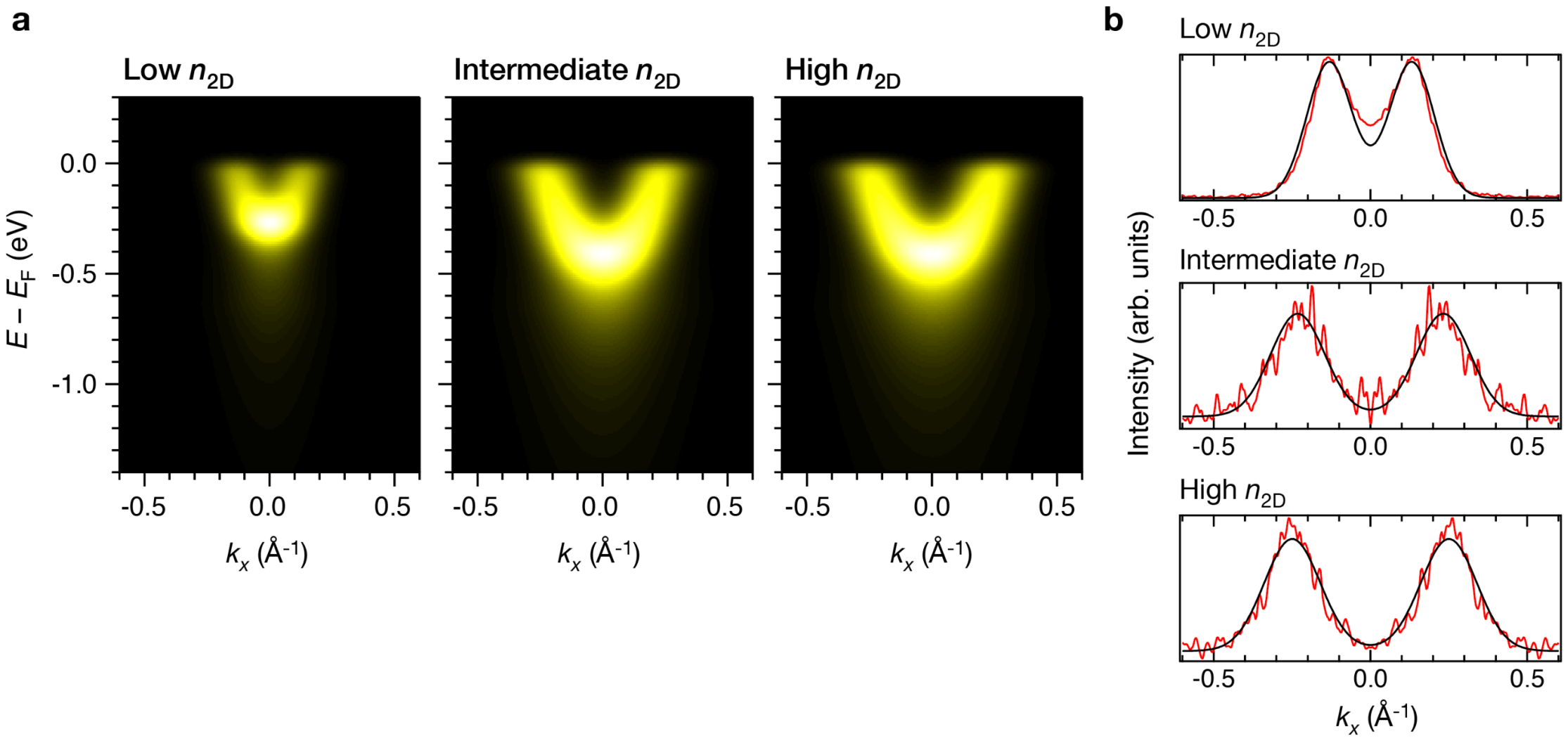

**Fig. S4 | Two-dimensional spectral-function fits with $k_F$ varied. a**, Calculated ARPES intensity maps for the low-, intermediate-, and high-$n_{2D}$ spectra, obtained from restricted-range two-dimensional spectral-function fits in which both the effective mass and $k_F$ were varied. The calculated intensities include the Fermi–Dirac distribution and Gaussian energy and momentum broadening. The fitted $k_F$ values are 0.139, 0.240, and 0.258 Å$^{-1}$, respectively. The maps extend beyond the energy windows used for fitting; the intensity at higher binding energies therefore represents an extrapolation of the fitted model. **b**, Momentum distribution curves integrated over a 20-meV-wide energy window centered at $E_F$. Red curves show the experimental spectra symmetrized about $k_x = 0$ Å$^{-1}$, as displayed in Fig. S2**a**, and black curves show the corresponding cuts through the calculated maps in **a**. The intensities were scaled for comparison.

**Table S1 | Effective masses and Fermi wave vectors obtained using different analysis procedures.** The effective masses are given in units of the free-electron mass, $m_e$. The values in parentheses in the final row are the $k_F$ values obtained from the fit, in Å$^{-1}$. The uncertainties for the MDC-based analyses represent formal one-standard-deviation fitting uncertainties. For the two-dimensional spectral-function fits, only the best-fit values are shown because the formal statistical uncertainties do not account for systematic uncertainties associated with the assumed spectral line shape, linewidth, background, momentum broadening, and pixel-to-pixel correlations. In both fits labeled "$k_F$ fixed," $k_F$ was set to 0.135, 0.224, and 0.252 Å$^{-1}$ for the low-, intermediate-, and high-$n_{2D}$ spectra, respectively, corresponding to $n_{2\mathrm{D}} = 2.9 \times 10^{13}$, $8.0 \times 10^{13}$, and $10.1 \times 10^{13}$ cm$^{-2}$ for a spin-degenerate circular Fermi surface. In the fit labeled "$k_F$ varied," $k_F$ was determined together with the effective mass. The robustness of the increase in $m^*$ was assessed by comparing the results of these fits with the independent MDC-based analyses. Both restricted-range fits, with $k_F$ either fixed or varied, used $E_B \leq 0.20$, 0.25, and 0.30 eV for the low-, intermediate-, and high-$n_{2D}$ spectra, respectively. The wide-range fit with fixed $k_F$ extended to $E_B = 1.0$ eV for the low- $n_{2D}$ spectrum and $E_B = 1.3$ eV for the intermediate- and high-$n_{2D}$ spectra.

| Analysis method | Low $n_{2D}$ | Intermediate $n_{2D}$ | High $n_{2D}$ |
|---|---|---|---|
| Parabolic fit | 0.23 ± 0.01 $m_e$ | 0.35 ± 0.01 $m_e$ | 0.48 ± 0.02 $m_e$ |
| Near-$E_F$ Fermi-velocity analysis | 0.20 ± 0.07 $m_e$ | 0.28 ± 0.06 $m_e$ | 0.50 ± 0.09 $m_e$ |
| 2D spectral-function fit, wide energy range ($k_F$ fixed) | 0.15 $m_e$ | 0.29 $m_e$ | 0.37 $m_e$ |
| 2D spectral-function fit, restricted energy range ($k_F$ fixed) | 0.22 $m_e$ | 0.34 $m_e$ | 0.50 $m_e$ |
| 2D spectral-function fit, restricted range ($k_F$ varied) | 0.24 $m_e$ (0.139 Å$^{-1}$) | 0.48 $m_e$ (0.240 Å$^{-1}$) | 0.55 $m_e$ (0.258 Å$^{-1}$) |

**Temporal stability and reproducibility of the ARPES-derived effective mass**

To assess whether slight surface charging or changes in the experimental configuration affect the extracted band dispersion, we examined both the temporal stability of the spectra under the lower-density condition, where the surface is expected to be more insulating, and the reproducibility of the dispersion under different measurement conditions.

Figure S5**a** shows a representative ARPES intensity map measured at $h\nu = 92$ eV in the very-low-density regime $n_{2D} < 2\times10^{12}$ cm$^{-2}$ with an acquisition time of approximately 2 min. Sequential angle-integrated spectra acquired over a total period of 20 min are overlaid in Fig. S5**b**. The spectra were normalized to their peak intensities to facilitate comparison of their energy positions and line shapes. The high-kinetic-energy cutoff exhibits a maximum displacement of only approximately 3 meV. Because each ARPES map used for the quantitative effective-mass analysis was acquired within 10 min, the time-dependent energy drift during an individual acquisition is bounded by this small value. Such a nearly rigid energy displacement is too small to account for the observed change in band curvature.

We further examined the reproducibility of the high-$n_{2D}$ dispersion using different beamlines and photon energies. Figures S5**c–e** show ARPES intensity maps measured at Photon Factory BL-2A with $h\nu = 92$ eV, and at NanoTerasu BL06U with $h\nu = 92$ eV and 102 eV, respectively. At each binding energy, the MDC was fitted using two Lorentzian peaks and a smooth background, and the resulting peak positions were fitted with a parabolic dispersion. The NanoTerasu measurements at both photon energies yielded $m^* = 0.48 \pm 0.02\ m_e$ for $n_{2D} = 1.0 \times 10^{14}$ cm$^{-2}$, whereas the Photon Factory measurement yielded $m^* = 0.47 \pm 0.02\ m_e$ for $n_{2D} = 9.9 \times 10^{13}$ cm$^{-2}$. The FWHM of each near-$E_F$ MDC peak was approximately 0.15 Å$^{-1}$ at NanoTerasu BL06U and 0.21 Å$^{-1}$ at Photon Factory BL-2A. The broader linewidth at Photon Factory may partly arise from spatial averaging associated with its larger beam spot of approximately $500 \times 200$ μm$^2$, compared with $20 \times 10$ μm$^2$ at NanoTerasu, although differences in instrumental resolution and local sample conditions may also contribute. Importantly, the broader linewidth does not produce a discernible difference in the extracted high-density band curvature or effective mass.

In addition, we examined a low-$n_{2D}$ state at the first $\bar{\Gamma}$ point. Figure S5**f** shows the ARPES dispersion measured at NanoTerasu BL06U using $h\nu =$ 92 eV. At $n_{2D} = 2.2 \times 10^{13}$ cm$^{-2}$, the parabolic fit yielded $m^* = 0.22 \pm 0.03$ $m_e$. This value is close to $m^* = 0.23 \pm 0.01$ $m_e$ obtained at $n_{2D} = 2.9 \times 10^{13}$ cm$^{-2}$ at the second $\bar{\Gamma}$ point, as shown in Figs. 3**c** and 3**e** of the main text. Under the same beamline, photon-energy, and $\bar{\Gamma}$-point configuration, $m^*$ increases from $0.22 \pm 0.03$ $m_e$ at $2.2 \times 10^{13}$ cm$^{-2}$ (Fig. S5**f**) to $0.48 \pm 0.02$ $m_e$ at $1.0 \times 10^{14}$ cm$^{-2}$ (Fig. S5**d**). These results show that the carrier-density-dependent increase in the effective mass is not attributable to the choice between the measured $\bar{\Gamma}$ points. The agreement between these independently acquired low-density spectra provides a check of the reproducibility of the extracted effective mass, although their slightly different carrier densities exclude using them as identical-state repeats for a statistical uncertainty estimate.

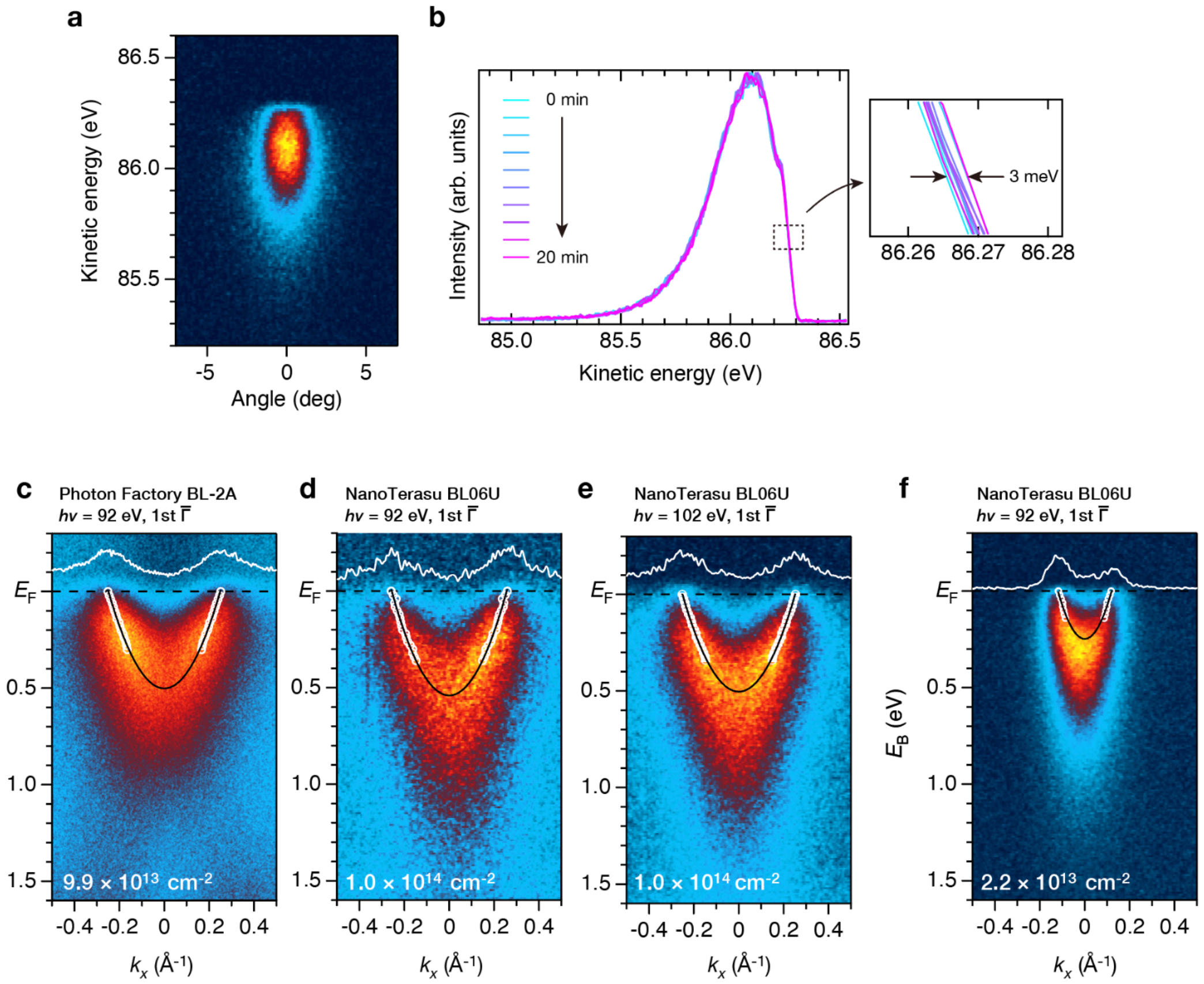

**Fig. S5 I Temporal stability and reproducibility of the ARPES-derived effective mass. a,** ARPES intensity map measured at $h\nu$ = 92 eV in the very-low-density regime $n_{2D} < 2\times10^{12}$ cm$^{-2}$. The acquisition time was approximately 2 min. **b,** Angle-integrated spectra acquired sequentially over a total period of 20 min and normalized to their peak intensities. The line color changes from cyan to magenta with increasing elapsed time. The inset shows an enlarged view of the high-kinetic-energy cutoff, whose maximum displacement is approximately 3 meV. **c-f,** ARPES intensity maps measured at the first $\bar{\Gamma}$ in the high-$n_{2D}$ regime (**c**-**e**) and in the low-$n_{2D}$ (**f**) under different experimental conditions: Photon Factory BL-2A at $h\nu$ = 92 eV (**c**), NanoTerasu BL06U at $h\nu$ = 92 eV (**d, f**), and NanoTerasu BL06U at $h\nu$ = 102 eV (**e**). All measurements were performed at 20 K using linear-horizontally polarized light. The white circles indicate the peak positions obtained by fitting each MDC with two Lorentzian components, and the black curves represent parabolic fits to the extracted dispersion. The white traces above the intensity maps are MDCs integrated over a 20-meV energy window centered at $E_F$ and are vertically offset for clarity. The corresponding carrier densities are indicated in each panel.

**Photon-energy dependence of the 2DES ARPES intensity**

Figure S6 shows the photon-energy dependence of the ARPES intensity from the 2DES formed on the electron-doped β-$Ga_2O_3$ surface. The background-subtracted normalized intensity, $(I_{k_\mathrm{F}} - I_\mathrm{BG})/I_\mathrm{BG}$, was plotted as a function of $k_z$, where $I_{k_\mathrm{F}}$ was obtained by integrating the spectral weight near the Fermi wave vector and $I_\mathrm{BG}$ was taken from a momentum region without 2DES spectral weight.

In the measured photon-energy range of $h\nu = 50 - 152$ eV, corresponding to the $k_z$ range shown in Fig. S6, the background-subtracted normalized intensity of the 2DES increases approximately monotonically overall with $k_z$. The photon-energy-dependent ARPES response of quantum-confined 2D states can, in principle, include the effect of the Fourier components of the confined out-of-plane wavefunction[5,6]. The widths of the corresponding intensity peaks reflect both the spatial confinement of the initial states and the finite photoelectron inelastic mean free path (IMFP)[5]. However, within the present photon-energy range, we do not observe a well-defined intensity maximum or periodic peak structure that would allow a quantitative determination of the confinement length. In the present VUV-ARPES range, IMFP-related broadening, non-free-electron final-state effects, and photoemission matrix elements can broaden or modify the intensity modulation. The present data do not allow these contributions to be separated and therefore do not provide an independent experimental determination of the confinement length.

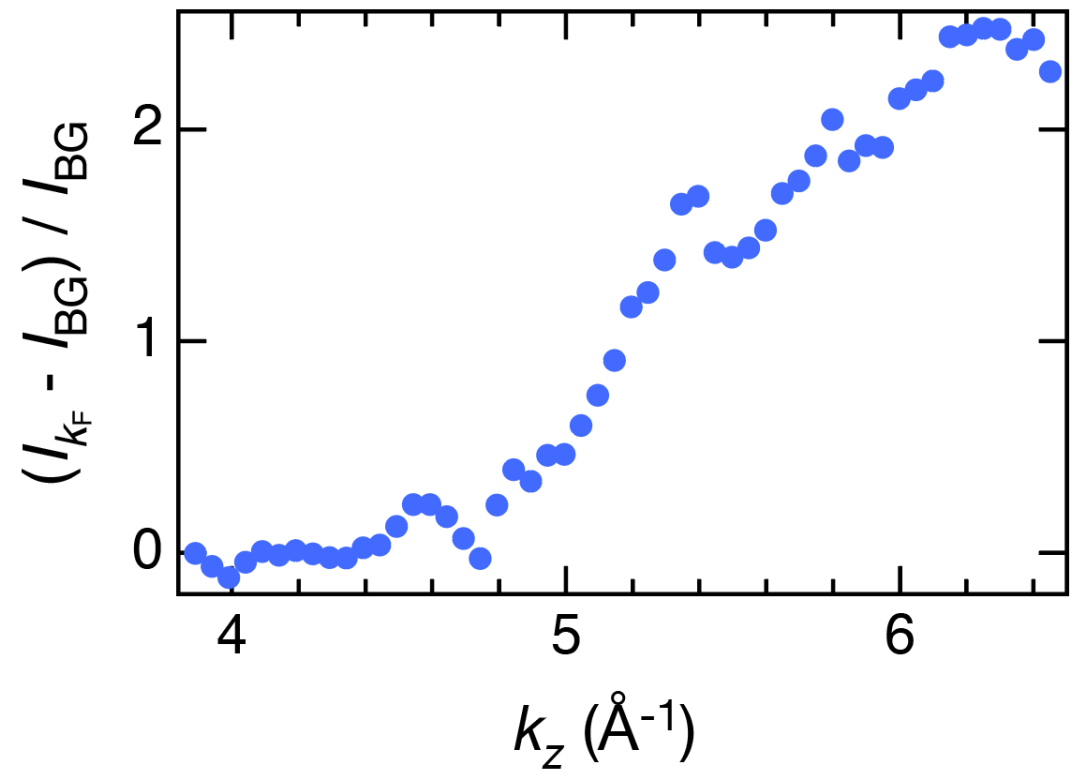

**Fig. S6 | Photon-energy dependence of the 2DES ARPES intensity.** Background-subtracted normalized intensity, $(I_{kF} - I_{BG})/I_{BG}$, plotted as a function of the out-of-plane momentum $k_z$. The data were obtained from the same photon-energy-dependent dataset as that shown in Fig. 1**f** of the main text, covering $h\nu$ = 50–152 eV. $I_{kF}$ was evaluated by integrating the spectral weight around the Fermi wave vector within an energy window of 50 meV and a momentum window of 0.05 Å$^{-1}$. $I_{BG}$ was obtained from a momentum region without 2DES spectral weight. The intensity exhibits an approximately monotonic overall increase with $k_z$, without a clearly resolved intensity maximum or periodic peak structure. Small features in the region $k_z$ = 5.2–6.0 Å$^{-1}$ may partly arise from core-level photoemission excited by higher-order light.

**Evaluation of the effective energy broadening and $k_F$ determination by the gradient method**

To evaluate the reliability of the $k_F$ determination, we first estimated the effective energy broadening from the Fermi edge of the Cs-doped metallic β-$Ga_2O_3$ surface. Figure S7 shows Fermi-edge spectra measured at $h\nu$ = 92 eV and 102 eV. The spectra were extracted from low-density data with $n_{2D} < 2 \times 10^{12}$ cm$^{-2}$ for which a clear Fermi edge and reliable fitting were obtained, by integrating the intensity around $k_x$ = 0 Å$^{-1}$ within a range of ±0.04 Å$^{-1}$. The $h\nu$ = 92 eV spectrum corresponds to the data shown in Fig. 3**a** of the main text. Fitting the spectra with a Fermi–Dirac distribution convoluted with a Gaussian function yielded effective energy broadenings (Gaussian FWHMs) of approximately 45 meV at *hv* = 92 eV and 50 meV at *hv* = 102 eV. These values are slightly larger than the reference value of 30 meV obtained from the Au Fermi edge, likely reflecting weak sample-related broadening. Importantly, the time-dependent shift of the spectral features was negligibly small during the measurements (Fig. S5**b**), indicating that the spectra were recorded under stable conditions.

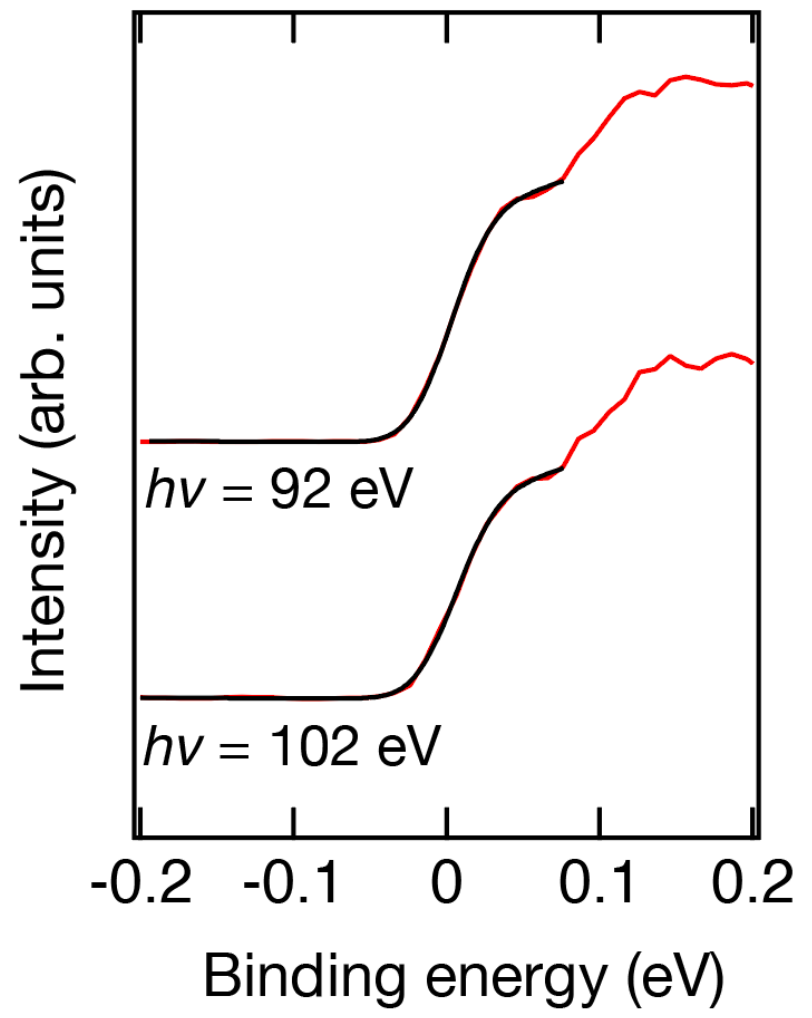

**Fig. S7 | Evaluation of the effective energy broadening.** Fermi-edge spectra of the Cs-doped metallic β-$Ga_2O_3$ surface measured at *hv* = 92 eV and 102 eV. The spectra were extracted from low-density data with $n_{2D} < 2 \times 10^{12}$ $cm^{-2}$, by integrating the intensity around $k_x = 0$ $Å^{-1}$ within a range of ±0.04 $Å^{-1}$. The *hv* = 92 eV spectrum corresponds to the data shown in Fig. 3**a** of the main text. The solid curves represent fits using a Fermi–Dirac distribution convoluted with a Gaussian function. The effective energy broadening is estimated to be approximately 45 meV for *hv* = 92 eV and 50 meV for *hv* = 102 eV.

Following the bandwidth-integrated gradient analysis of Straub *et al.*[7] and the comparison of full-bandwidth and near-$E_F$ gradients by Strocov *et al.*[8], we integrated the ARPES intensity over the occupied 2DES band to obtain $w(k_x)$ and estimated $k_F$ from the maxima of $|dw(k_x)/dk_x|$. Figure S8 shows $w(k_x)$ and $|dw(k_x)/dk_x|$ for the spectra shown in Fig. 3**a**–**d** of the main text. The energy integration windows were $-0.10 \leq E_B \leq 0.15$ eV for Fig. 3**a**, $-0.10 \leq E_B \leq 0.40$ eV for Figs. 3**b** and 3**c**, and $-0.10 \leq E_B \leq 0.60$ eV for Fig. 3**d**.

For the data with $n_{2D} \gtrsim 7.6 \times 10^{12}$ $cm^{-2}$, the $k_F$ values obtained by the gradient method are larger than those obtained from the MDC peak positions, by approximately 0.01–0.03 $Å^{-1}$. Because $w(k_x)$ contains energy- and momentum-broadened spectral weight together with a nonzero background, the maximum of $|dw(k_x)/dk_x|$ can be displaced outward relative to the band-crossing $k_F$. For the lowest-density data, the occupied bandwidth is much smaller, with the band bottom estimated to be around $E_B \sim 100$ meV. Because this energy scale is only a few times larger than the effective energy broadening, the bandwidth-integrated intensity does not form a well-defined plateau-like momentum distribution. Consequently, the gradient maximum can be affected by the broadened spectral tail and background, resulting in a larger apparent value of $k_F \sim 0.07$ $Å^{-1}$.

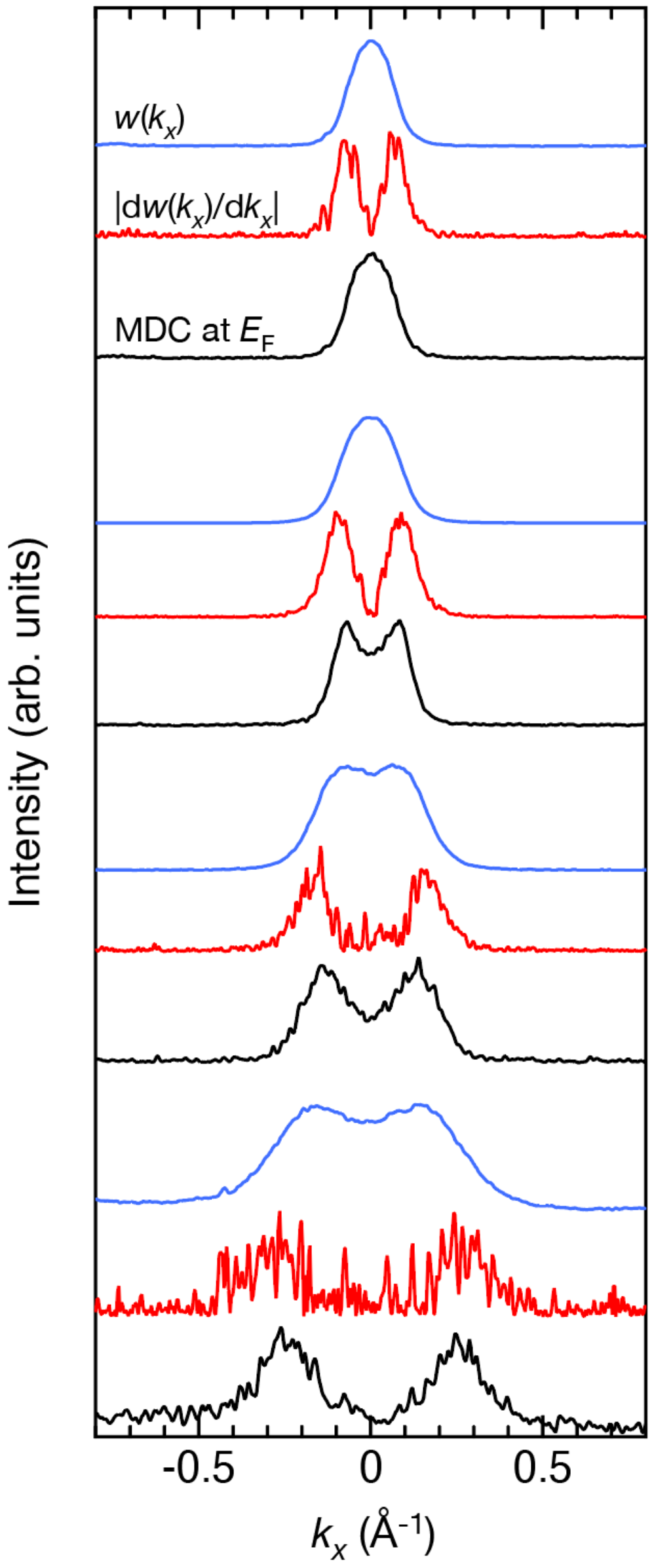

**Fig. S8 | Determination of $k_F$ using the gradient method.** Bandwidth-integrated ARPES intensity $w(k_x)$, its absolute derivative $|dw(k_x)/dk_x|$ and MDCs integrated over a 20-meV energy window centered at $E_F$ for the 2DES bands shown in Fig. 3**a**–**d** of the main text, displayed from top to bottom. The blue and red curves represent $w(k_x)$ and $|dw(k_x)/dk_x|$, respectively. The energy integration windows were $-0.10 \leq E_B \leq 0.15$ eV for Fig. 3**a**, $-0.10 \leq E_B \leq 0.40$ eV for Figs. 3**b** and 3**c**, and $-0.10 \leq E_B \leq 0.60$ eV for Fig. 3**d**. For the data with $n_{2D} \gtrsim 7.6 \times 10^{12}$ cm$^{-2}$, the $k_F$ values obtained from the gradient method are larger than those obtained from the MDC peak positions by

approximately 0.01–0.03 Å$^{-1}$. For the lowest-density data, the gradient method gives a larger value of $k_F \sim 0.07$ Å$^{-1}$, reflecting the larger uncertainty associated with the shallow occupied bandwidth and weak spectral weight.

**Angular analysis of the near-$E_F$ momentum contour**

To quantitatively evaluate the in-plane symmetry of the near-$E_F$ electronic states, a momentum map was constructed by integrating the photoemission intensity over an energy window of $E_F \pm$ 50 meV, corresponding to a total integration width of 100 meV. It should be noted that this energy-integrated map is not a strict Fermi-surface map, and the analysis below is therefore used to compare the momentum-contour shape along different in-plane directions.

MDCs were extracted from the energy-integrated map at angular intervals of 10° from 0° to 170°, where 0° is defined as the direction parallel to $k_x$. Each MDC was fitted using two Lorentzian peak components and a smooth background, as shown in Fig. S9**a**. The peak positions on the positive- and negative-momentum sides were assigned to the corresponding directions over the full angular range from 0° to 350°. The complete set of radial peak positions was empirically represented using a rotated ellipse. The formal one-standard-deviation uncertainties returned by the individual MDC fits were used as weights in the nonlinear least-squares fitting. The elliptical function was used only to interpolate the measured near-$E_F$ contour and to compare the radial peak positions along the $k_x$ and $k_y$ directions. Confidence intervals were evaluated using 5,000 Monte Carlo trials. In each trial, the individual MDC peak positions were independently varied according to Gaussian distributions defined by their formal fitting uncertainties, and the resulting contour was refitted. The radial peak positions evaluated along the $k_x$ and $k_y$ directions were $k_{\mathrm{peak,x}}$ = 0.2319 Å$^{-1}$ and $k_{\mathrm{peak,y}}$ = 0.2313 Å$^{-1}$, respectively (Fig. S9**b**). Their difference was $k_{\mathrm{peak,y}} - k_{\mathrm{peak,x}}$ = -0.0006 Å$^{-1}$ with a 95% confidence interval from -0.0073 to 0.0062 Å$^{-1}$. Because this confidence interval includes zero, no statistically significant difference between the $k_x$ and $k_y$ directions was resolved at the 95% confidence level. It should be noted that the absolute radial peak positions are smaller than the $k_F$ value ($k_F$ = 0.25 Å$^{-1}$) obtained from the band crossings (Fig. 1**e**) because the intensity was integrated over a wider energy window around $E_F$.

The confidence interval above reflects the formal statistical uncertainties of the MDC fits. Additional uncertainties associated with the initial fitting parameters, fitting range, background function, and peak line shape could broaden the confidence interval and shift its central value; however, they are not expected to change the qualitative conclusion that no statistically significant difference is resolved within the present experimental precision.

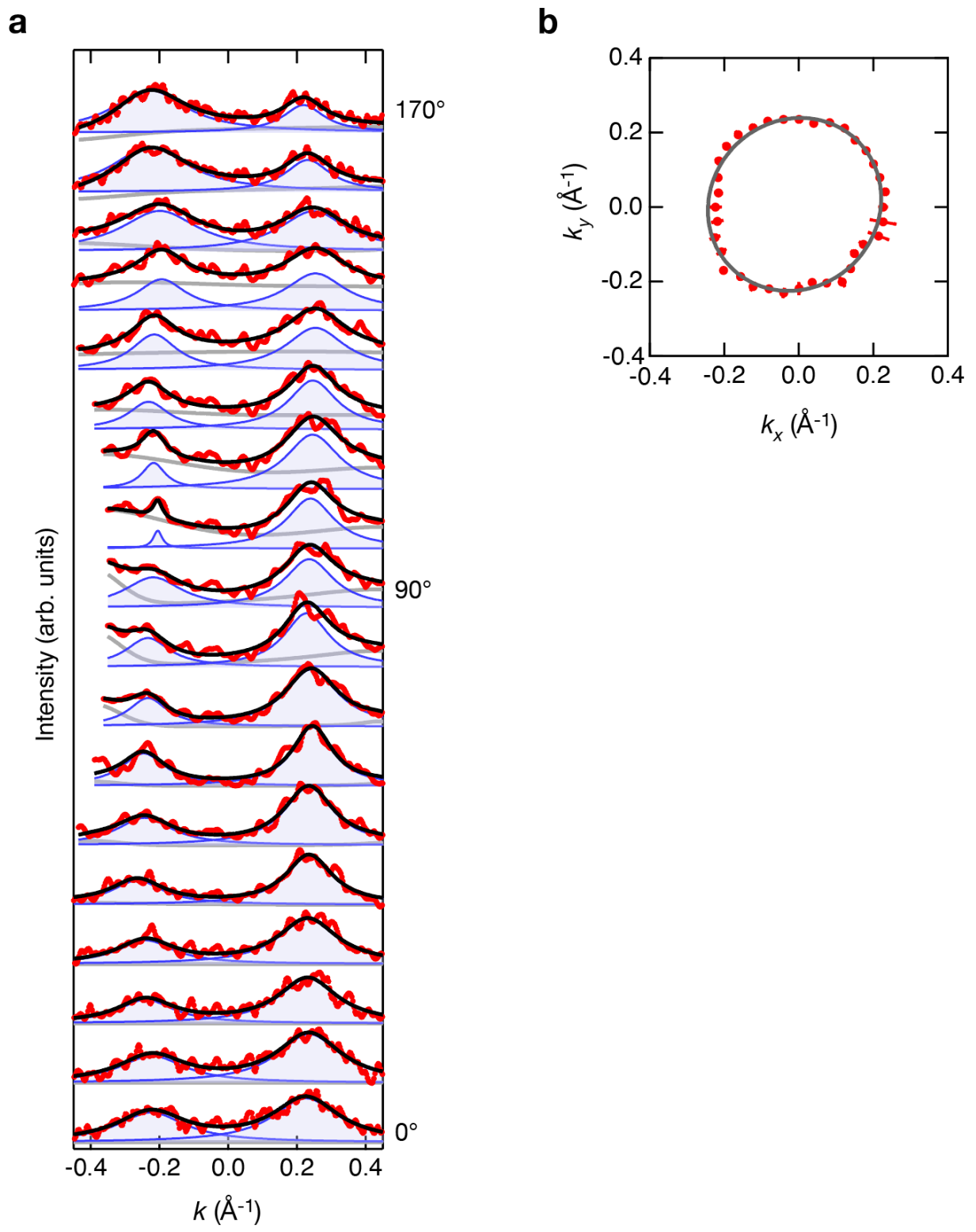

**Fig. S9 | Angular analysis of the energy-integrated near-$E_F$ momentum contour. a**, MDCs extracted at angular intervals of 10° from a momentum map obtained by integrating the photoemission intensity over $E_F \pm 50$ meV. The 0° direction is parallel to $k_x$. Red circles show the experimental MDCs after subtraction of the constant baseline. Gray curves represent the fitted background contributions, blue shaded curves represent the two Lorentzian peak components, and black curves show the total fits. The spectra are vertically offset for clarity. **b**, Radial peak positions obtained from the MDC fits and plotted in the $k_x$ - $k_y$ plane. The radial error bars represent the individual formal one-standard-deviation uncertainties of the fitted MDC peak positions. The black curve is an empirical fit of the complete contour using a rotated ellipse with a free center position.

## Calculated and experimental band structures

Figure S10 presents the Brillouin zone (**a**) and the bulk band structures of β-$Ga_2O_3$ computed using density-functional theory (DFT; **b**) and the single-shot $G_0W_0$ approximation (**c**), together with experimental valence-band ARPES spectra (**d**,**e**). The resulting fundamental band gaps are 2.0 eV (DFT) and 3.8 eV ($G_0W_0$).

The experimental spectra were measured on the Cs-free β-$Ga_2O_3$ surface at photon energies of 96 and 118 eV, sampling regions near the X–Γ–X and $X_1$–Y–$X_1$ directions, respectively. Multiple dispersive valence-band features are resolved. For these spectra, $E_{F,ref}$ denotes a nominal energy reference set 4.7 eV above the valence-band maximum, whereas the calculated band structures are referenced to the conduction-band minimum. The experimental path assignments are approximate because $k_z$ varies with binding energy and emission angle.

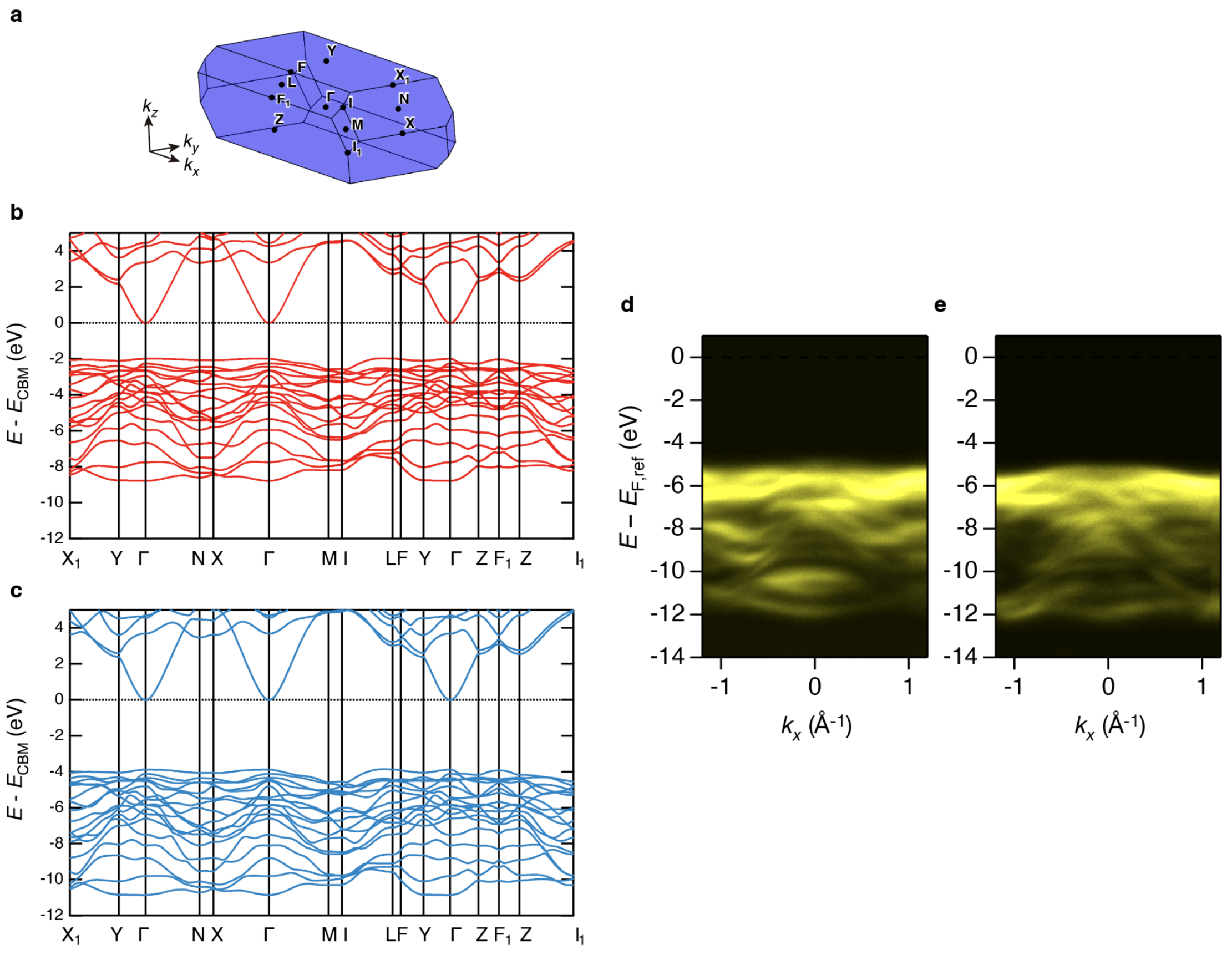

**Fig. S10 | Calculated and experimental band structures of the bulk β-$Ga_2O_3$. a**, The Brillouin zone with high-symmetry points labeled following the notation of Peelaers and Van de Walle[9]. **b**,**c**, Calculated band structure of bulk β-$Ga_2O_3$ crystal by DFT (**b**) and $G_0W_0$ (**c**), where 0 eV is set to the conduction band minimum (CBM). **d**,**e**, Experimental valence-band ARPES intensity maps measured on the Cs-free β-$Ga_2O_3$ surface at photon energies of 96 eV (**d**) and 118 eV (**e**), sampling regions near the X–Γ–X and $X_1$–Y–$X_1$ directions, respectively. In **d**,**e**, $E_{F,ref}$ denotes a nominal energy reference set 4.7 eV above the valence-band maximum. The path assignments refer approximately to the normal-emission $k_z$ values at this reference energy; $k_z$ varies with binding energy and emission angle.

**Additional calculated results**

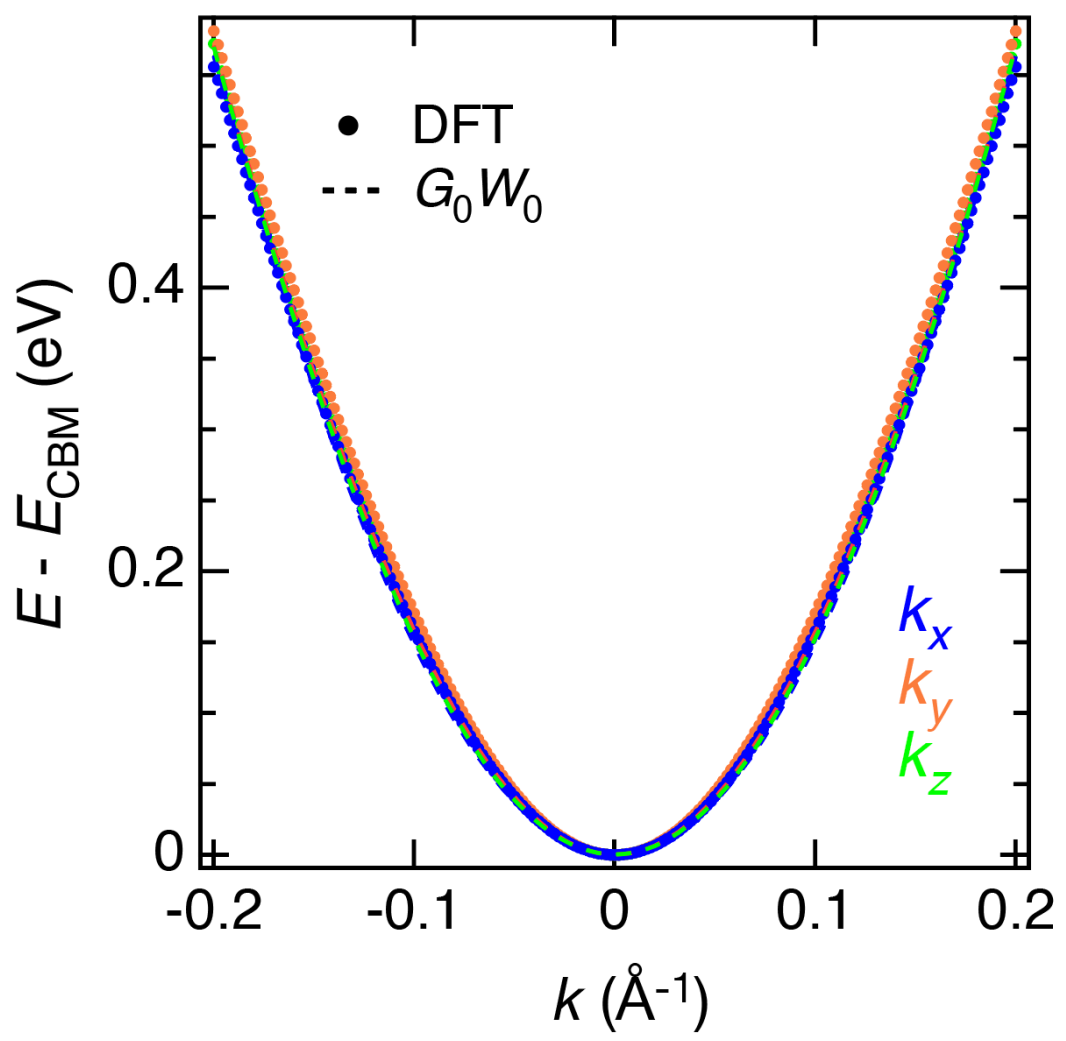

**Fig. S11 l Calculated conduction bands of the bulk β-$Ga_2O_3$.** Calculated conduction-band dispersions near the CBM along the $k_x$, $k_y$, and $k_z$ directions. Calculation results for both DFT and $G_0W_0$ are shown. Here, 0 Å$^{-1}$ corresponds to the Γ point of the Brillouin zone.

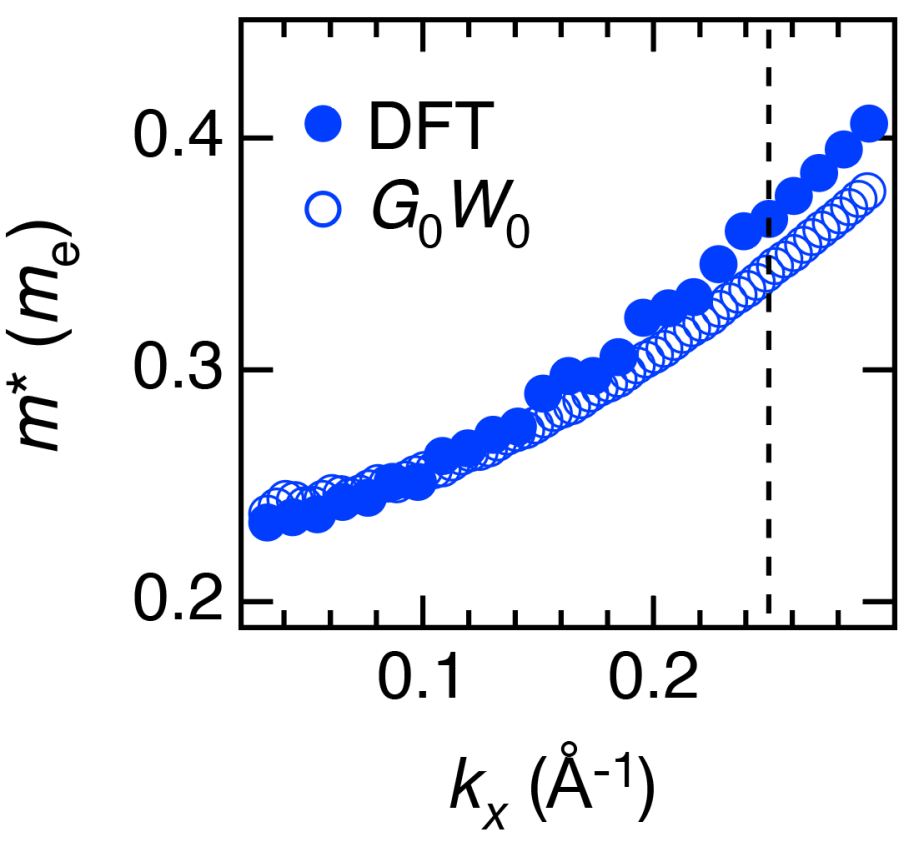

**Fig. S12 I Calculated effective masses of the bulk β-$Ga_2O_3$.** $k_x$ dependence of the effective-mass fit. $k_x$ width was set to 0.016 Å$^{-1}$ for all the fittings. The dashed line at $k_x$ = 0.25 Å$^{-1}$ indicates the $k_F$ at the highest carrier density ($n_{2D} = 1.0 \times 10^{14}$ cm$^{-2}$) observed in the ARPES measurements (Fig. 3**d** of the main text).

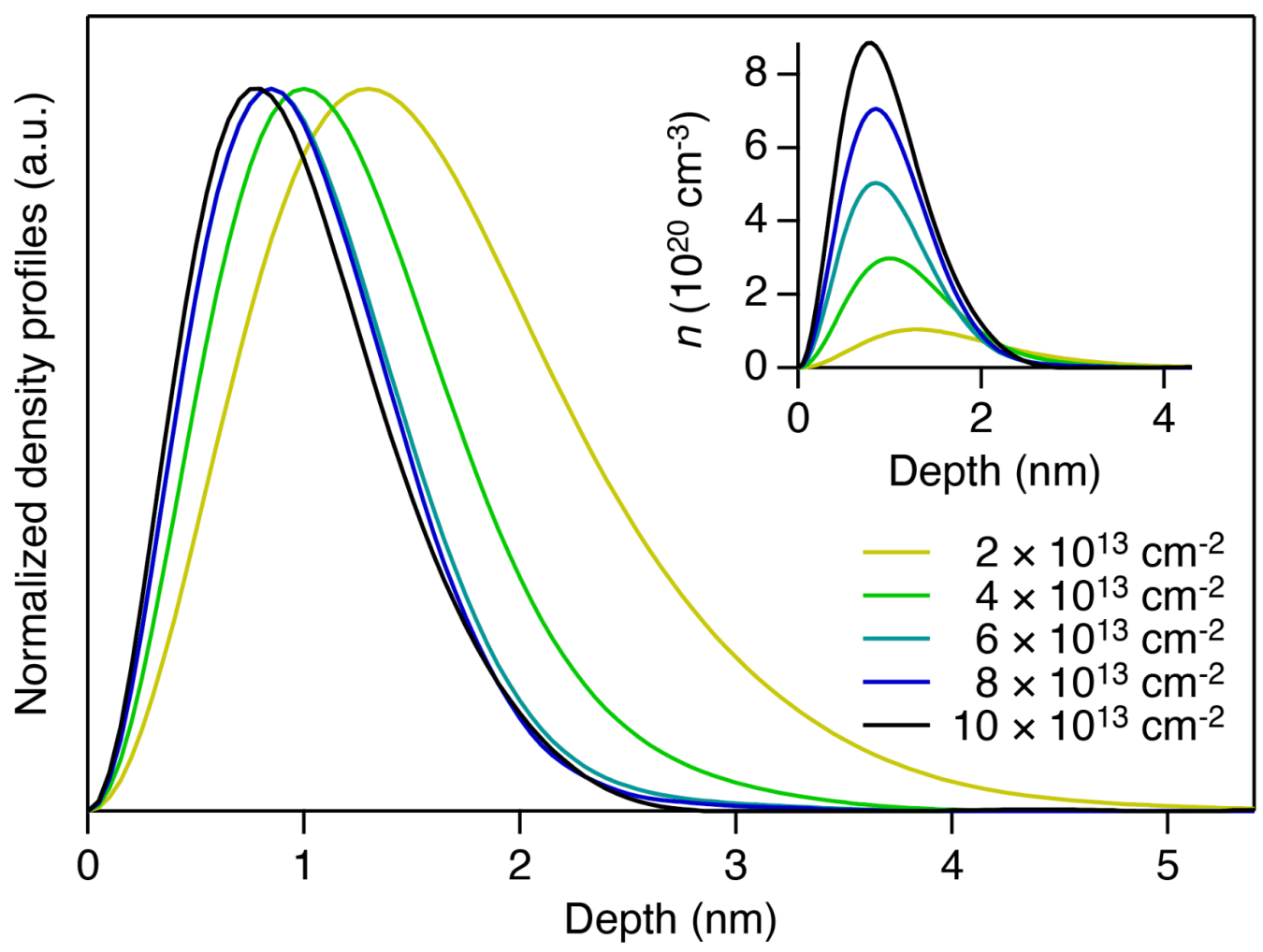

**Fig. S13 I Evolution of density profiles with $n_{2D}$.** Carrier-density profiles of the first subband obtained by self-consistently solving the Poisson–Schrödinger equations at various $n_{2D}$ values. To enable a straightforward comparison of changes in the density-profile shape, the isotropic effective mass was fixed to 0.23 $m_e$, close to the value at low sheet-carrier density (Fig. 3**g**). Therefore, the profile at $10 \times 10^{13}$ cm$^{-2}$ differs from that shown in Fig. 2**c**, where 0.48 $m_e$ was used.

**Supplementary References**